\documentclass[useAMS,usenatbib]{mn2e}
\usepackage[pdftex]{graphicx}
\usepackage{ifpdf}

\def\kmsmpc{km s$^{-1}$ Mpc$^{-1}$}
\def\ergshz{ergs s$^{-1}$ Hz$^{-1}$}
\def\halpha{\ifmmode {\rm H{\alpha}} \else $\rm H{\alpha}$\fi}
\def\hbeta{\ifmmode {\rm H{\beta}} \else $\rm H{\beta}$\fi}
\def\oiii{[O\,{\sc iii}] $\lambda\lambda$4959,5007}

\def\oiiib{[O\,{\sc iii}] $\lambda$5007}
\newcommand{\msun}{\,{\rm M_\odot}}
\DeclareRobustCommand{\gtrsim}{%
	\mathrel{\hskip-.5em\begin{array}{c}>\\[-8pt]\sim\end{array}\hskip-.5em}}
\newcommand{\ltapprox}{\raisebox{-0.5ex}{$\,\stackrel{<}{\scriptstyle\sim}\,$}}

\newcommand{\vrot}{\hbox{$V_{rot}$}}
\newcommand{\whalf}{\hbox{$S_{0.5}$}}
\newcommand{\sigtwod}{\hbox{$\sigma_{mean}$}}
\newcommand{\sigoned}{\hbox{$\sigma_{1D}$}}

\title[2D KINEMATICS AND PHYSICAL PROPERTIES OF $z  \sim 3$ STAR-FORMING GALAXIES]{2D KINEMATICS AND PHYSICAL PROPERTIES OF $z  \sim 3$ STAR-FORMING GALAXIES}
\author[M. Lemoine-Busserolle et al.]{M. Lemoine-Busserolle$^{1}$\thanks{Current address: Gemini Observatory, 670 N. A\'Ohoku Place, Hilo, 96720 Hawaii;  E-mail:
mbusserolle@gemini.edu} A. Bunker$^{1}$, F. Lamareille$^{2}$ and M. Kissler-Patig$^{3}$  \\
$^{1}$Oxford Physics, University of Oxford, Keble Road, Oxford, OX1\,3RH, UK\\
$^{2}$Laboratoire d'Astrophysique de Toulouse-Tarbes, Universit\'e de Toulouse, CNRS, 14 av. E. Belin, 31400 Toulouse, France\\
$^{3}$ESO, Karl-Schwarzschild-Str.2, 85748 Garching - Germany} 
\begin{document}

\date{Accepted ??. Received 2009 February 19}

\pagerange{\pageref{firstpage}--\pageref{lastpage}} \pubyear{2009}

\maketitle

\label{firstpage}

\begin{abstract}
	We present results from a study of the kinematic structure of star-forming galaxies at redshift $z
\sim 3$ selected in the VVDS, using integral-field spectroscopy of rest-frame optical nebular emission
lines, in combination with rest-frame UV spectroscopy, ground-based optical/near-IR and Spitzer
photometry. We also constrain the underlying stellar populations to address the evolutionary status of
these galaxies. We infer the kinematic properties of four galaxies: VVDS-20298666, VVDS-020297772,
VVDS-20463884 and VVDS-20335183 with redshifts $z=3.2917$, $3.2878$, $3.2776$, and $3.7062$,
respectively. While VVDS-20463884 presents an irregular velocity field with a peak in the local
velocity dispersion of the galaxy shifted from the centre of the galaxy, VVDS-20298666 has a
well-resolved gradient in velocity over a distance of $\sim$4.5 kpc with a peak-to-peak amplitude of $v=$91
km s$^{-1}$. We discovered that the nearby galaxy, VVDS-020297772 (which shows traces of AGN activity), is in
fact a companion at a similar redshift with a projected separated of 12\,kpc. In contrast, the
velocity field of VVDS-020335183 seems more consistent with a merger on a rotating disk. However, all of
the objects have a high local velocity dispersion ($\sigma\sim$ 60-70 km s$^{-1}$), which gives $v/\sigma
\ltapprox 1$. It is unlikely that these galaxies are dynamically cold rotating disk of ionized
gas.
\end{abstract}

\begin{table*}
\caption{SINFONI Observations}
	\resizebox{18cm}{!} {
		\begin{tabular}{ccccccccc}
		\hline
 Galaxy &   VVDS-ID  &    RA (J2000) &        DEC (J2000) &          $z$ (1) &  Exp.\ Time (2)  &     Seeing (3)  &    Airmass (4) & Dates of Obs.\ (U.T.) \\
		\hline
 VVDS-3884 & 020463884 &    02 27 01.23 & -04 21 04.4 &      3.2780 &    3.5 Hrs &     0\farcs38      &   1.28         & 07 Sept.\ 2005 \\
VVDS-8666 & 020298666 &    02 26 18.32 & -04 20 32.7 &     3.2805 &      5 Hrs &       0\farcs54     &     1.12       & 06,08 Sept.\ 2005 \\
 VVDS-7772 &  020297772 & 02 26 18.23 & -04 20 34.1 & ---$^{\dagger}$ &      5 Hrs &       0\farcs54     &     1.12       & 06,08 Sept.\ 2005 \\
 VVDS-5183 &  020335183 &    02 25 33.71 & -04 15 41.4 &     3.6993 &      6 Hrs &       0\farcs38     &     1.07       & 13,15 Nov.\ 2006 \\
		\hline
		\end{tabular}
	}
$^{\dagger}$ This nearby galaxy to VVDS-8666 was discovered in our IFU data cube
to share the same redshift. The original photometric redshift from VVDS/CFHT-LS was $z_{phot}=1.747$
with broad minimum in $\chi ^2$. It was not observed spectroscopically with VIMOS as part of the VVDS survey.
(1) redshift estimated from rest-frame UV spectrum obtained with VIMOS.
(2) Total observing time usable for each target (900s for each individual exposures).
(3) median seeing estimated from the PSF stars.
(4) average airmass of all the exposures.
\label{runs_sum}
\end{table*}
\section{Introduction}
\label{Intro}
To understand the assembly of stars in galaxies, it is crucial to characterize the star formation rate and the total
stellar mass already formed as a function of the
dynamical mass (the mass of the halo including dark matter). Determining this relation at intermediate and high redshifts allows us to explore whether we are seeing the initial build-up of stellar mass in a fully-formed dark halo, or if the stellar mass to total mass ratio is fairly constant compared to lower-redshift samples (as might be the case if much of the stellar mass formed very early, or dark haloes were assembled through dry mergers triggering little further star formation). The redshifts  range $z \sim 2 - 4$  is therefore of great interest as massive galaxies  are thought to undergo their most active period of star formation (e.g, \citealt{2006ApJ...651..142H}).

In the past, deep multi-wavelength surveys (e.g., GOODS -- \citealt{2004ApJ...600L..93G}; Hubble Deep Field -- \citealt{1996AJ....112.1335W}; DEEP2
-- \citealt{2007ApJ...665..265F}) have highlighted the large diversity in morphology and stellar populations among high redshift galaxies. Many
studies using UV--IR photometry and long-slit spectroscopy, mainly focusing on the Lyman-break galaxy (LBG) populations (e.g.
\citealt{2003ApJ...591..101E,2004ApJ...612..122E,2006ApJ...644..813E}) and complemented by various studies using lensed objects (e.g.,
\citealt{2003A&A...397..839L,2000ApJ...531...95B}), have helped to understand the global characteristics of $z \sim 2 - 4$ galaxies. However,
with the limited information available from long-slit spectroscopy it is difficult to answer keys questions on the dynamical nature of high
redshift galaxies. It is plausible that some may be perturbed by bursts of star formation and merging to lie outside the low-redshift
morphological classification scheme (the familiar Hubble sequence). To properly explore whether this is the case, or instead if most
star-forming galaxies have stable disk kinematics, requires studying the resolved kinematic structure through ``3D" spectroscopy with Integral
Field Units (IFUs). The recent advent of high resolution IFUs on 8m-class telescopes has enabled the study of high-redshift galaxies on scales
of a few kiloparsecs in seeing-limited conditions (e.g.\ \citealt{2004AN....325..139B,2004MNRAS.354L..19S}) and even down to sub-kiloparsec scales using adaptive optics and the magnification due to gravitational lensing (e.g.,
\citealt{2008Natur.455..775S}). Very recent studies using Integral Field Spectroscopy have started to give an insight into the kinematics
properties and physical phenomena at play in high redshift galaxies, such as merging and galactic winds associated with strong star formation
\citep{2006Natur.442..786G,2007ApJ...669..929L,2009ApJ...697.2057L,2007ApJ...658...78W,w2,2006ApJ...645.1062F,2007ApJ...671..303B,2008ApJ...687...59G,2008A&A...479...67N,2008A&A...488...99V}.
There is some controversy from this early work about the properties of the dynamical structure of these distant galaxies (particularly the
proportion of rotationally supported gaseous disks versus objects kinematically dominated by high velocity dispersions). However, there is a
consensus from the work so far that at $z>2$ galaxies typically exhibit higher velocity dispersions (relative to ordered rotation) compared
with lower-redshift samples.

In this paper, we present results on the kinematic properties of a sample of four galaxies at $3<z<4$ selected in the VVDS Deep data ($I<24$;
\citealt{2005A&A...439..845L}), using IFU $K$-band spectroscopy with VLT/SINFONI. These spectra are sensitive to the rest-frame optical, in
particular the nebular emission lines from gas ionized by hot stars in regions of recent star formation. Hence we can determine the current
star formation rate, and the spatial distribution of star formation, as well as measuring the kinematics of the gas. We are able to study the
stellar populations through CFHT photometry from both the VVDS and CFHTLS surveys, and also imaging with the Spitzer Space Telescope using IRAC
and MIPS, complemented by UV rest-frame spectroscopy from VLT/VIMOS.  The results presented here are part of a study to construct 
samples representative of the global intermediate and high-$z$ population by selecting galaxies only on the basis of their magnitude ($I_{AB}
<24.75$). Two companion papers, Lemoine-Busserolle et al.\ (2009b, in revision) and Queyrel et al.\ (2009, in revision) are devoted to the
study of intermediate redshift galaxies ($1< z  < 1.5$) observed with VLT/SINFONI during the same observing runs.

In Section \ref{Obs} we describe our sample, observational strategy and the data reductions techniques. In Section \ref{sprectro_prop} we
address the nature of the stellar population of the galaxies, using broad-band photometry to constrain properties such as stellar mass, age and
star formation rate. In Section \ref{kine_prop} we investigate the kinematic structure and dynamical properties inferred from the integral
field spectroscopy. Finally in Section \ref{summary} we summarize our results and explore the nature of these LBG galaxies at $z=3-4$. We
compare our findings with previous studies of kinematic properties of $z \sim 2 - 3$ galaxies.

We assume a cosmology with $\Omega_{0} = 0.3$, $\Lambda = 0.7$ and $H_{0}$ = 70 \kmsmpc\ throughout, and all magnitudes are on the $AB$ system \citep{1983ApJ...266..713O}.
\begin{figure*}
\begin{center}
\resizebox{2.0\columnwidth}{!}{\includegraphics{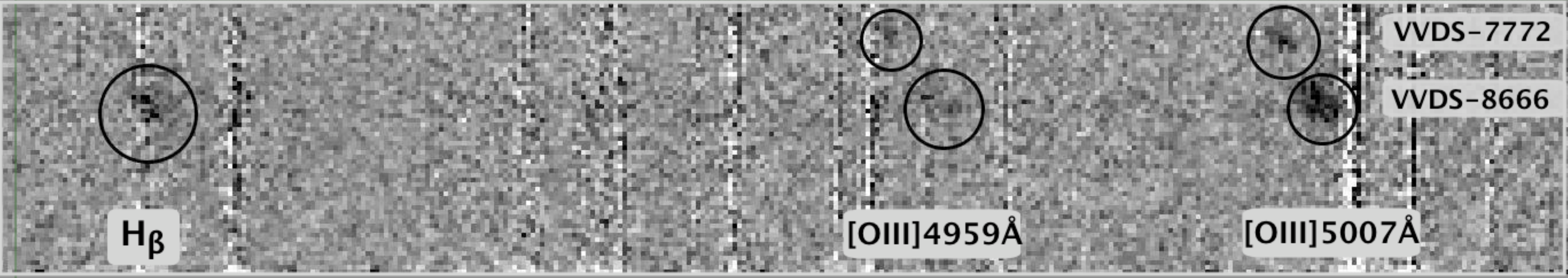}}
\end{center}
\caption{Two-dimensional spectrum of the nebular emission lines of VVDS-8666 and VVDS-7772. The $x$-axis is wavelength (increasing to the right) and spans the range $\lambda=20800-21620$\,\AA . The $y$-axis is spatial distance, with a length of $6\farcs0$ displayed.}
\label{comp2D}
\end{figure*}
\section{Observations and Data Reduction} 
\label{Obs} 
\subsection{Sample Selection} 
\label{sample}
Our IFU observations targetted a sample of 3 galaxies selected in the 02h ``deep'' ($17.5 \leq I_{AB} \leq 24.0$) field of the VIMOS VLT Deep
Survey (VVDS; \citealt{2005A&A...439..845L}). The $z>3$ objects presented here were selected on the basis of redshifts derived from the
VLT/VIMOS rest-UV spectra, such that the expected wavelengths of the rest-optical H$\beta$/[O\,III] emission lines would be clear of
bright OH sky lines in the near-IR.
\subsection{SINFONI Observational Strategy} 
The near-IR spectroscopic observations of the sample were acquired with the 3D-spectrograph SINFONI (SINgle Faint Object Near-IR Investigation --
\citet{2003SPIE.4841.1548E}) at ESO-VLT during two 4-nights runs, on September 5-8, 2005 (ESO run 75.A-0318) and on November 12-15, 2006 (ESO
run 78.A-0177). We used the largest-field mode of SINFONI, to enable good sky subtraction without the need to do offset sky observations. This
mode has a spaxel scale of $0\farcs125\times 0\farcs25$ pixels (with coarser sampling in $y$), leading to a field-of-view of 8\arcsec\ $\times$
8\arcsec. Our observations used the $K$-grism, spanning $1.94-2.46\,\mu$m with 2.45\,\AA\ pixels, and are detailed in Table~\ref{runs_sum}.
Unresolved sky lines had a measured width of $\approx$ 7\,\AA\ (FWHM), indicating a spectral resolution $R_S=\lambda/\Delta\lambda_{\rm FWHM}
\approx 3300$ (or 90\,km/s velocity resolution). Conditions were photometric and the median seeing for each objects is in Table~\ref{runs_sum}.
Each target was acquired through a blind offset from a nearby bright star. Individual integrations were 900\,s, and we read out the array in
multiple non-destructive read mode to reduce the readout noise to $\sim 7\,e^{-}$ and become background-limited. Between readouts the telescope
was nodded by $\approx 4$\arcsec, positioning the galaxy in opposite corners within the 8\arcsec\ $\times$ 8\arcsec\ SINFONI field-of-view. Two
sets of observations located around the upper corner (A) and the lower corner (B) were obtained per object. This observational procedure allows
background subtraction by using frames contiguous in time, but with the galaxy in different locations. Moreover, the target was never located
exactly at the same position on the detector, a minimal sub-dithering of 0.3\arcsec\ was required in order to minimize instrumental artifacts
when the individual observations are aligned and combined together. The target was reacquired from the offset star every hour. The blind offset
stars were also used to monitor the PSF. The total on-source integration times are listed in Table~\ref{runs_sum}. SINFONI was used in its
seeing-limited mode for two galaxies (VVDS-020298666 in September 2005, and VVDS-020335183 in November 2006 -- hereafter abbreviated to VVDS-8666 and VVDS-5183), and in closed-loop NGS-AO
(Natural Guide Star -- Adaptive Optics) mode for the observations of VVDS-020463884 (hereafter called VVDS-3884) on 07 September 2005, which had a suitably bright $K=10.7$
nearby star $27\farcs4$ away. The Strehl ratio achieved at 2.2\,$\mu$m was $15\%$.
\subsection{SINFONI Data Reduction and Flux Calibration} 
Data reduction has been performed with the ESO-SINFONI pipeline (version 1.7.1, \citealt{2007astro.ph..1297M}), IRAF and custom IDL scripts. 
The individual 900s exposures were run through the pipeline, which flat-fields the frames using an internal lamp, corrects for hot pixels using dark frames,
wavelength calibrates each individual observation from the spectra of reference arc lamps, and then reconstructs and spatially registers the
three-dimensional datacubes (with the sky background still present) for each exposure after applying a distortion correction. Our
observing strategy was implemented to maximize on-source integration and therefore no offset blank-sky frame was obtained. As the field of view used with SINFONI is larger than the spatial extent of the galaxies, we were able to obtain a clean sky
background subtraction by stepping the galaxies between two well-separated points within the field of view, using the approach and IDL script presented in \cite{2007MNRAS.375.1099D}. Sky background subtraction was done
directly on the cube, using the individual cubes at position B as a sky for the individual cube at position A and vice versa. After sky
subtraction, individual cubes were aligned in the spatial direction by relying on the telescope offsets from the nearby bright star (used as a
reference for the blind acquisition) and then combined together, rejecting remaining cosmic ray strikes to produce the final reduced cube for
each object.

A flux calibration is required in order to derived absolute parameters (e.g.\ star formation rate) from the uncalibrated flux measured in
emission lines. Each science observation has been accompanied by the observation of a telluric standard star at similar airmass of B spectral
type and magnitude $K=6-7$. Integration times were $2-6$\,s for these standard star observations, and the cubes were reduced in a similar way
to our science exposures. We extracted 1D spectra of the stars (summing all the flux within an aperture of diameter 5
resolution elements) which were used both to perform the flux calibration of our galaxy cubes, and to correct them from telluric absorption
lines. We assume that the standard stars are well fitted by a pure blackbody curve (a good approximation for these B stars). Starting from the
blackbody temperature of the standard star and its magnitude in $K$-band, we calibrated the sensitivity function using IRAF. The galaxy cubes
were then divided by the calibration curve appropriate for that observation to produce a flux calibrated spectrum, corrected for atmospheric
absorption. At the centre of the $K$-band, the conversion was approximately $1\,{\rm count/s}=2\times 10^{-17}\,{\rm erg\,cm^{-2}\,s^{-1}}$.

We measured the noise in each data cube by taking a slice in wavelength close to the position of the line emission. We then determined the RMS counts in blank 
areas uncontaminated by emission from the target galaxy
in the region where  all the dithered pointings overlapped. In computing the flux uncertainty in an emission line from the extracted spectrum, we
multiplied the measured pixel-to-pixel noise in the data cube by $\sqrt{\pi r_{spat}^2d_{\lambda}}$, where $r_{spat}$ is the radius in pixels
of the extraction aperture, and $d_{\lambda}$ is the wavelength extent (again in pixels) over which the line flux is measured. For the galaxies
VVDS-5183, VVDS-8666 and VVDS-3884 we used spatial apertures of $r_{spat}=8$\,pixels ($r=1\farcs0$), and for the compact companion VVDS-02029772 (hereafter abreviated to VVDS-7772) we used a smaller aperture
of $r_{spat}=4$\,pixels ($r=0\farcs5$). In all cases, we used an extraction of width 8 pixels in wavelength (20\,\AA\ or 300\,km/s), which was
$\gtrsim$ FWHM of the lines. The typical $1\,\sigma$ noise was $0.2-0.3\times 10^{-17}\,{\rm erg\,cm^{-2}\,s^{-1}}$ for our extraction aperture.

The blind offset stars were reduced in the same way 
than the telluric standards. We produced collapsed 2D
images using the SINFONI pipeline to allow to  the extent
of the PSF to be estimated from the offset stars.
\begin{table*}
	\caption{Nebular emission lines measurements from SINFONI 1D spectra}
	\resizebox{18cm}{!} {
	\begin{tabular}{ccccccccccc}
		\hline
        Galaxy & $\lambda$(H$\beta$) (1) & $z_{H\beta}$ (2) & $\Delta\lambda$(H$\beta$) (3) & f(H$\beta$) (4) &  $\lambda$(\oiiib ) (1) & $z_{\rm [O{\scriptsize~III]}}$ (2) & $\Delta\lambda$(\oiiib ) (3) & $f$(\oiiib ) (4) &  $f$(\oiiib ) \\
        & /\,\AA\ & & FWHM / \AA & /\,$10^{-17}$\,erg\,cm$^{-2}$\,s$^{-1}$ & /\,\AA\ & & FWHM / \AA & /\,$10^{-17}$erg\,cm$^{-2}$\,s$^{-1}$ & /\,$f$(H$\beta$) \\
		\hline
  VVDS-3884 &  $20797.81$ &  $3.2782$  &   $21.52$       & $5.67\pm0.28$     & 21417.50 &  3.2776  &     $22.21$ &     $16.17\pm0.34$ &    $2.85$ \\

  VVDS-8666 &  $20866.36$  & $3.2923$  &   $18.43$   &    $4.44\pm0.20$       &     21487.68   & 3.2917    &     $14.96$ &     $7.10\pm0.20$ &     $1.59$  \\

  VVDS-7772 &   ---  & ---   &  ---      &      $<0.3$ ($2\,\sigma$)      &   21468.79   & 3.2878   & $11.27$ &     $2.58\pm0.14$ &     $>9$ ($2\,\sigma$) \\

  VVDS-5183 & 22876.63 & 3.7058 & $16.60$  & $7.40\pm0.30$    &  23563.85 & 3.7063  & $18.82 $        &     $24.10\pm0.50$       &   $3.26$ \\
	\hline
	\end{tabular} 
}
(1) Wavelength of nebular emission line.
(2) redshift estimated from rest-frame optical nebular emission line.
(3) FWHM of emission line.
(4) Emission line flux with uncertainties.  Non-detections are recorded as  $2\,\sigma$ upper limits.
\label{tab:IR_lines}
\end{table*}
\section{Spectrophotometry properties} 
\label{sprectro_prop} 
\subsection{VVDS-7772: A companion galaxy of VVDS-8666 at $z=3.29$}
\begin{figure}
\begin{center}
\resizebox{0.9\columnwidth}{!}{\includegraphics{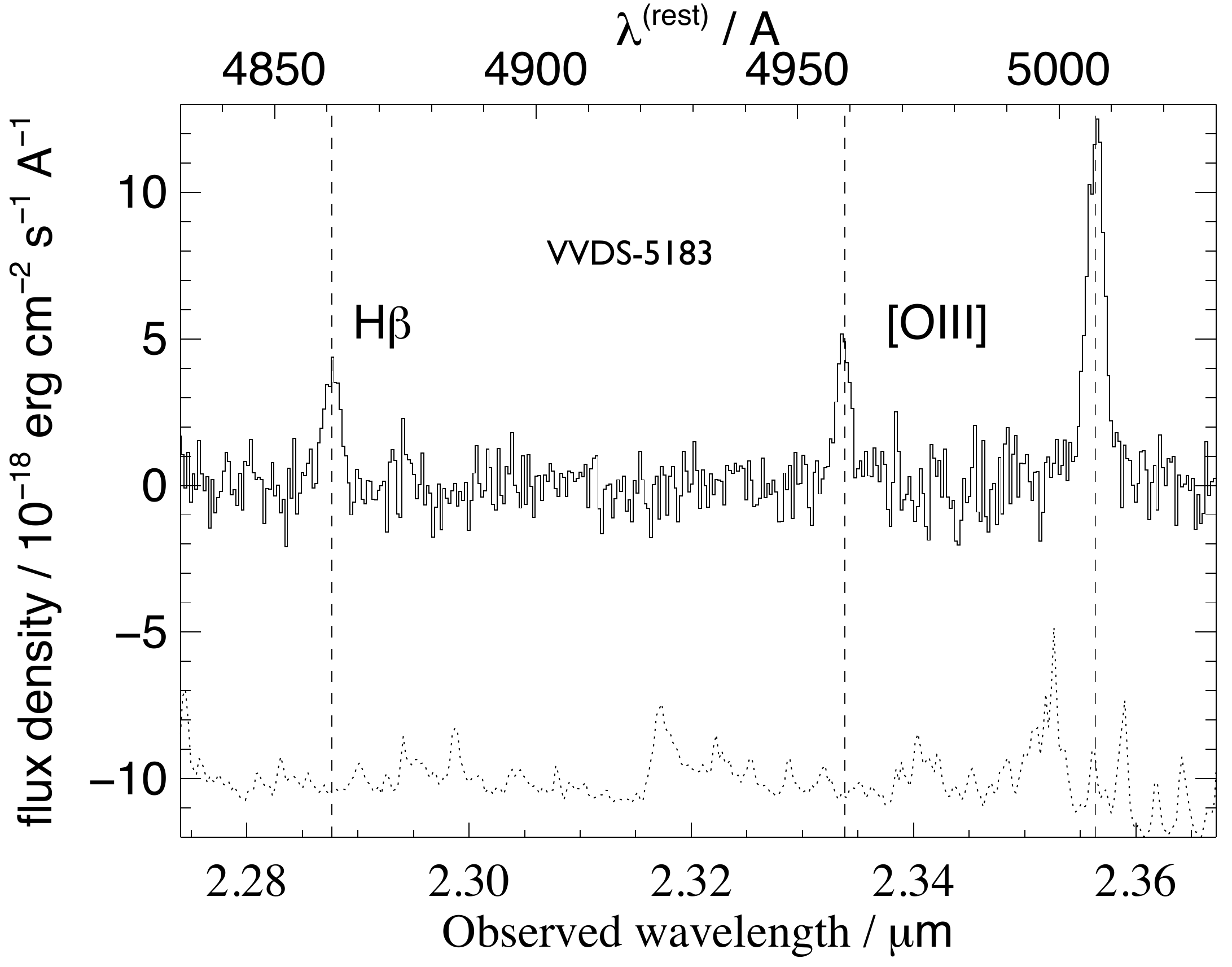}}
\resizebox{0.9\columnwidth}{!}{\includegraphics{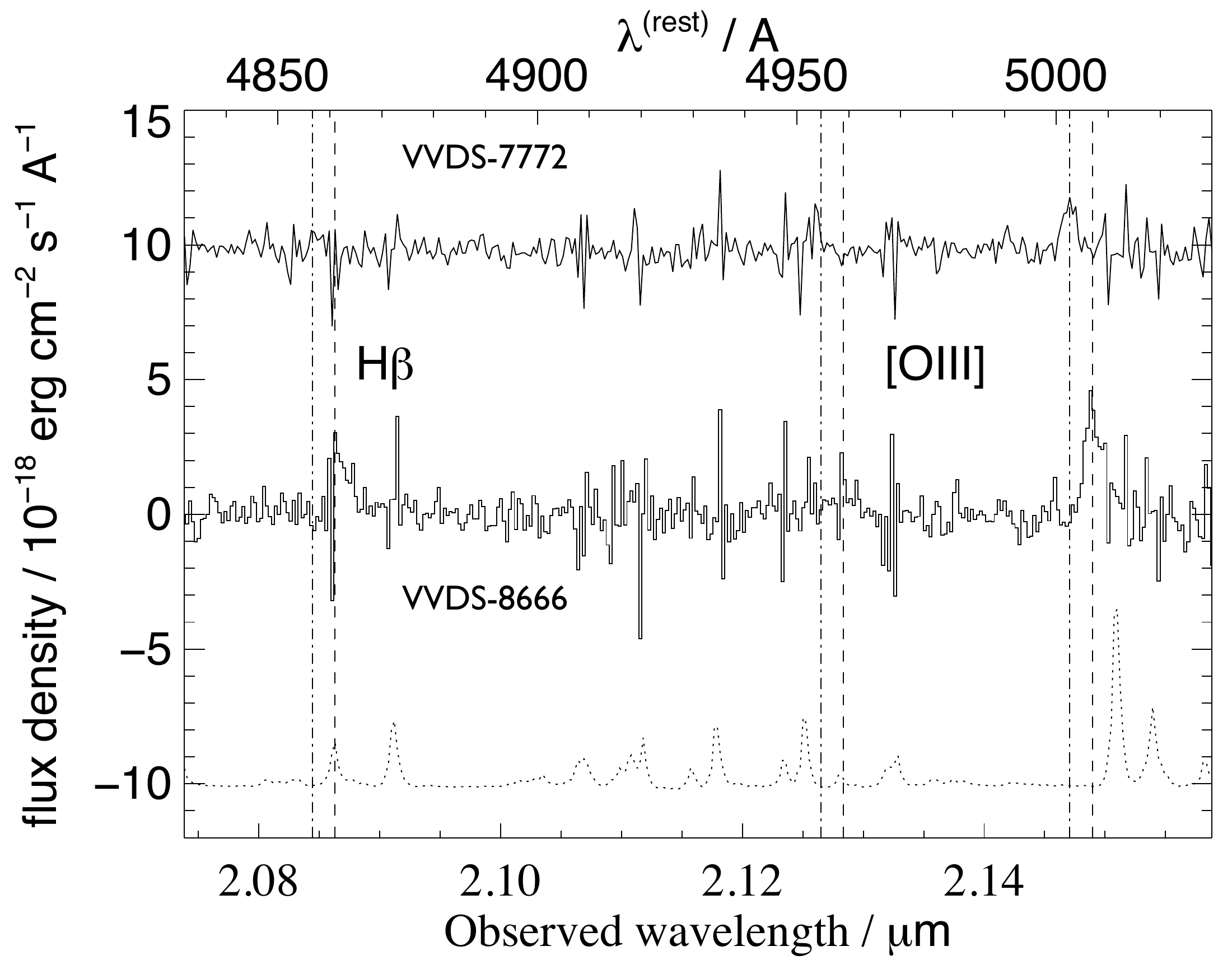}}
\resizebox{0.9\columnwidth}{!}{\includegraphics{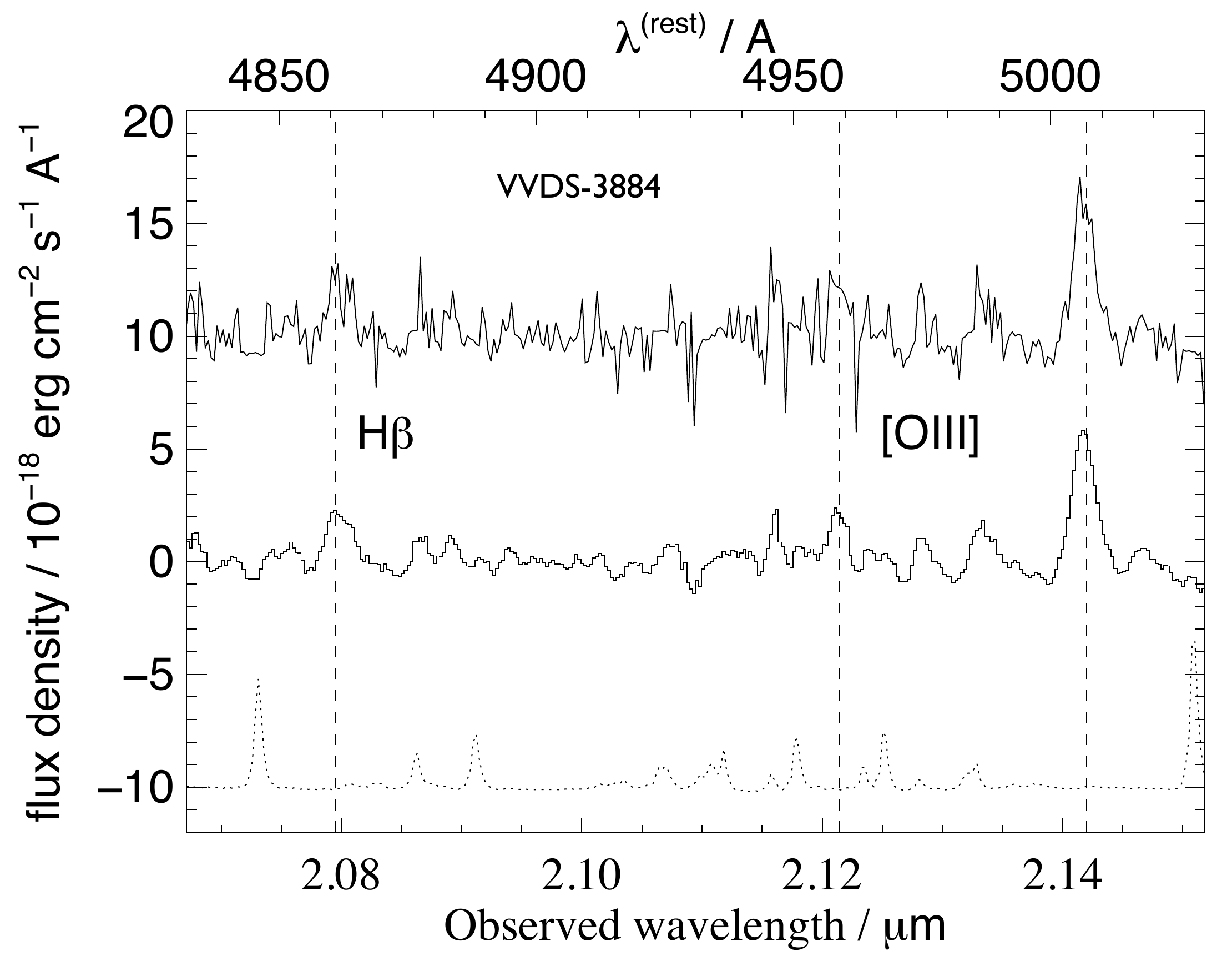}}
\end{center}
\caption{SINFONI integrated 1D spectra of the nebular emission lines, \hbeta\ to \oiii (indicated by vertical dashed lines). At the bottom of
each panel, a scaled version of the sky spectrum is shown (dotted line). Galaxy VVDS-5183 is shown in the top plot, with VVDS-8666 in
the middle panel (with the spectrum of the nearby companion, VVDS-7772, shown offset by $10^{-17}$ above); the bottom panel shows the spectrum of
VVDS-3884 (smoothed with a boxcar of 15 pixels; the unsmoothed data is shown offset above).}
\label{IR_1D}
\end{figure}

Our IFU spectroscopy of VVDS-8666 reveals a companion galaxy with [O\,III] emission at a similar redshift. The [O\,III] emission of the
companion is blue-shifted by $307\pm34$\,km/s relative to the targetted galaxy (see Figs.~\ref{comp2D} \& \ref{IR_1D}). The companion is included in the
VVDS photometric catalog as object VVDS-020297772 (hereafter VVDS-7772), with $I_{AB}=25.0$, and is undetected in $U$-band. The photometric redshift is only
$z_{phot}=1.7478$, very different from our spectroscopic redshift of $z_{spec}=3.2878$, which may be due to an unusual SED (perhaps
contaminated by the presence of an AGN, see Section~\ref{metallicity}).

The spatial separation of the galaxies is $1\farcs6\pm 0\farcs2$ from the line emissions in the 3D IFU cube, which corresponds to a projected
distance of 12\,kpc at $z=3.29$ (the nominal separation in the broad-band images in worse seeing is $1\farcs 9$). If we assume that these two
galaxies are gravitationally bound, and the velocity offset is due to orbital motion and not outflows etc., the enclosed mass is given by
$v^2r/G$, and we can set a lower bound on the orbital radius, $r$, by using the projected separation (12\,kpc) and also the resolved component
of the orbital velocity along the line of sight (307\,km\,s$^{-1}$). This yields $M_{\rm system}>2.6\times 10^{11}\,M_{\odot}$.
\begin{figure}
\begin{center}
\resizebox{1.0\columnwidth}{!}{\includegraphics{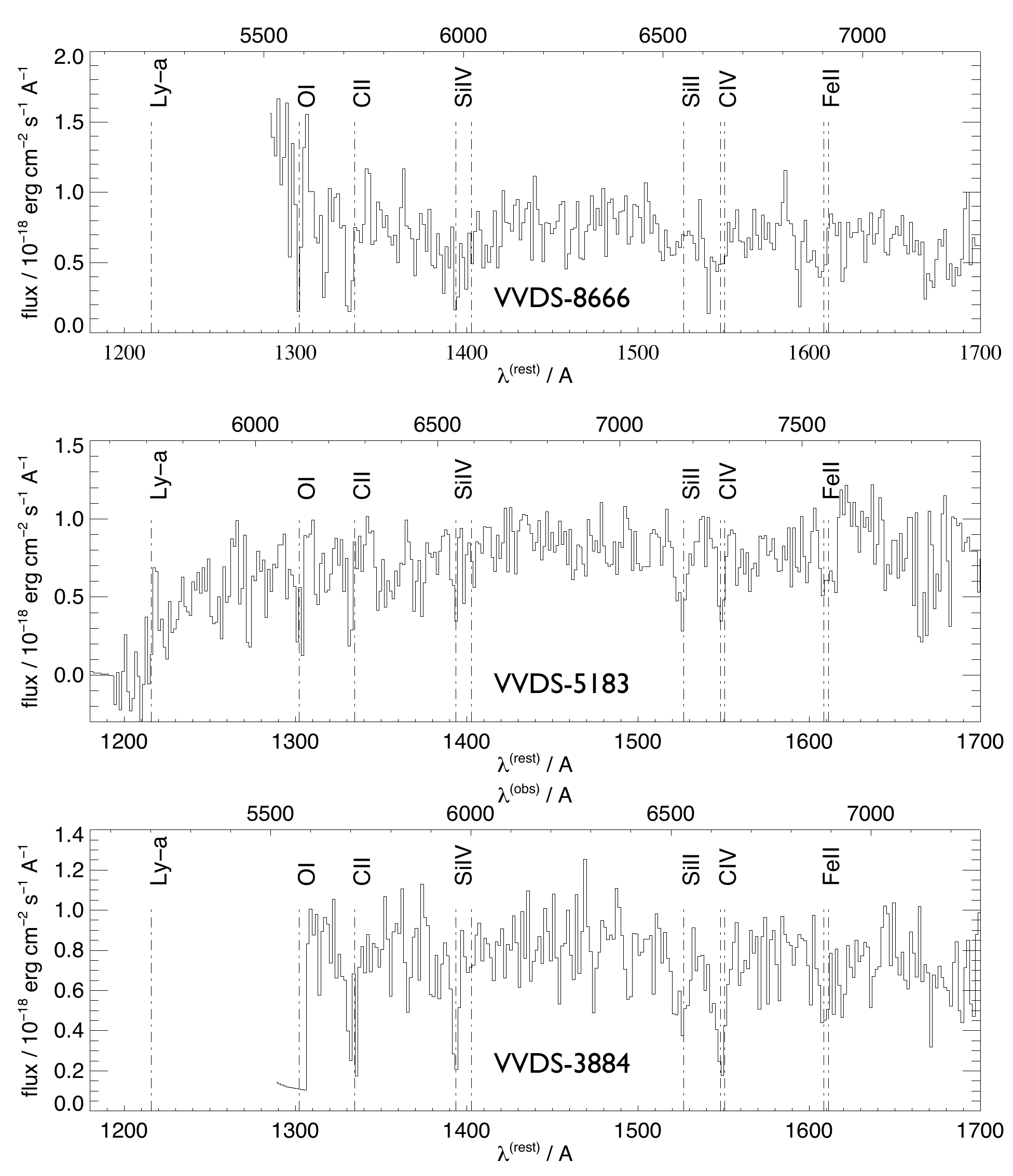}}
\end{center}
\caption{One-dimensional UV-rest frame spectra of VVDS-3884,VVDS-5183 and VVDS-8666. Vertical dotted lines indicate the wavelengths of the strongest spectral features}
\label{UV_1D}
\end{figure}
\subsection{Rest-Frame UV Spectroscopy}
Optical spectra have been obtained with the VIMOS instrument installed at ESO/VLT as part of the VVDS survey
(Fig.~\ref{UV_1D}). The spectral coverage is $5600$\,\AA\,$<\lambda<9250$\,\AA\
which corresponds approximately to rest-frame UV wavelengths $1300$\,\AA\,$<\lambda<2150$\,\AA\ for the $z\approx 3.3$ galaxies, and $1190$\,\AA\,$<\lambda<1970$\,\AA\ for VVDS-5183 at $z\approx 3.7$ (which encompasses Lyman-$\alpha$\,1216\,\AA\ in this high redshift case, but not in the other galaxies). The resolution is $R_s\approx230$. Of the three galaxies, VVDS-8666 has a relatively featureless spectrum
with only O{\scriptsize~I}\,1302\,\AA / C{\scriptsize~II}\,1304\,\AA\ absorption prominent. VVDS-3884 has strong absorption features from the high-ionization interstellar medium (ISM) lines C{\scriptsize~IV}\,1548,1550\,\AA\ and Si{\scriptsize~IV}\,1394,1504\,\AA\ (with rest-frame equivalent widths of $W_0=4.5$\,\AA\ and $3.0$ respectively), as well as weaker C{\scriptsize~II}\,1334\,\AA\ and Si{\scriptsize~II}\,1526\,\AA\ absorption. The high-ionization lines are blueshifted with respect to the [O{\scriptsize~III}] redshift by $\approx 100$\,km\,s$^{-1}$, with the low ionization lines blueshifted by $\approx 300$\,km\,s$^{-1}$. There is a hint of He{\scriptsize~II}\,1640\,\AA\ emission in the 1D spectrum. In contrast, VVDS-5183 has more prominent low-ionization ISM absorption lines than the high-ionization C{\scriptsize~IV} \& Si{\scriptsize~IV}. All the ISM lines are blueshifted by $\sim 500$\,km\,s$^{-1}$ relative to the [O{\scriptsize~III}] redshift. This galaxy exhibits weak Lyman-$\alpha$ emission
  of $6\times 10^{-18}\,{\rm ergs\,cm^{-2}\,s^{-1}}$ that is redshifted by 200\,km\,s$^{-1}$ relative to [O{\scriptsize~III}],
with a rest-frame equivalent width of $W_0\approx -7$\,\AA .
\subsection{Rest-Frame Optical Spectroscopy and Metallicity Estimates from Nebular Lines}
\label{metallicity}
Integrated rest-frame optical 1D spectra were extracted from the SINFONI data cube for each object. From these we measured line fluxes and velocity
widths of the integrated line emission (see Fig.~\ref{IR_1D}). We can use the [O{\scriptsize~III}]/H$\beta$ ratio to investigate any possible contribution
of an AGN to the nebular spectrum. We follow the classification proposed by \citet{2009A&A...495...53L}.
This classification is based on the distribution of galaxies of known type in the 2dFGRS survey, with respect to their [O{\scriptsize~III}]/H$\beta$ ratio.
The classification as a star-forming galaxy is secure for log([O{\scriptsize~III}]/H$\beta$) lower than 0.4, and the classification as an AGN is secure for 
 log([O{\scriptsize~III}]/H$\beta$) greater than 0.7. It is impossible to classify with certainty galaxies lying in the region between these two limits,
 but we know that this region is dominated by star-forming galaxies ($\approx$ 60\% for log([O{\scriptsize~III}]/H$\beta$) lower than 0.6).
 With a log([O{\scriptsize~III}]/H$\beta$) of 0.46 and
0.51, respectively, VVDS-3884 and VVDS-5183 are candidate star-forming galaxies. The nebular spectra of these two galaxies are
probably produced in hot H{\scriptsize~II} star-forming regions. However, there is also a chance ($\approx$ 40\%) that they are instead dominated by an AGN (i.e.\
Seyfert 2 galaxies). With log([O{\scriptsize~III}]/H$\beta$) of 0.20, VVDS-8666 is without doubt a star-forming galaxy. Finally, the H$\beta$ line from
the companion (VVDS-7772) is undetected, unlike in VVDS-8666, and we derive an upper-limit for H$\beta$ emission by collapsing the
data cube around the wavelength inferred from companion redshift from [O{\scriptsize~III}], and performing aperture photometry and background subtraction on the
2D line map. We quote a $2\,\sigma$ upper limit from the local background noise in the 3D data cube. We can thus compute a lower limit for
log([O{\scriptsize~III}]/H$\beta$) which is 0.93. With such a large [O{\scriptsize~III}]/H$\beta$ ratio, VVDS-7772 is probably dominated by an AGN.

Gas-phase oxygen abundances may be also estimated from the [O{\scriptsize~III}]/H$\beta$ ratio (see Table~\ref{tab:IR_lines}) using the relation provided by \citet{2006ApJ...652..257L},
which was calibrated on \citet{2004ApJ...613..898T} metallicities in the SDSS DR4 sample. 
We find 12+log(O/H) of $8.57\pm0.02$, $8.66\pm0.02$, and $8.56\pm0.02$ for VVDS-3884, VVDS-8666, and VVDS-5183, respectively. 

\begin{figure}
\begin{center}
\resizebox{0.9\columnwidth}{!}{\includegraphics{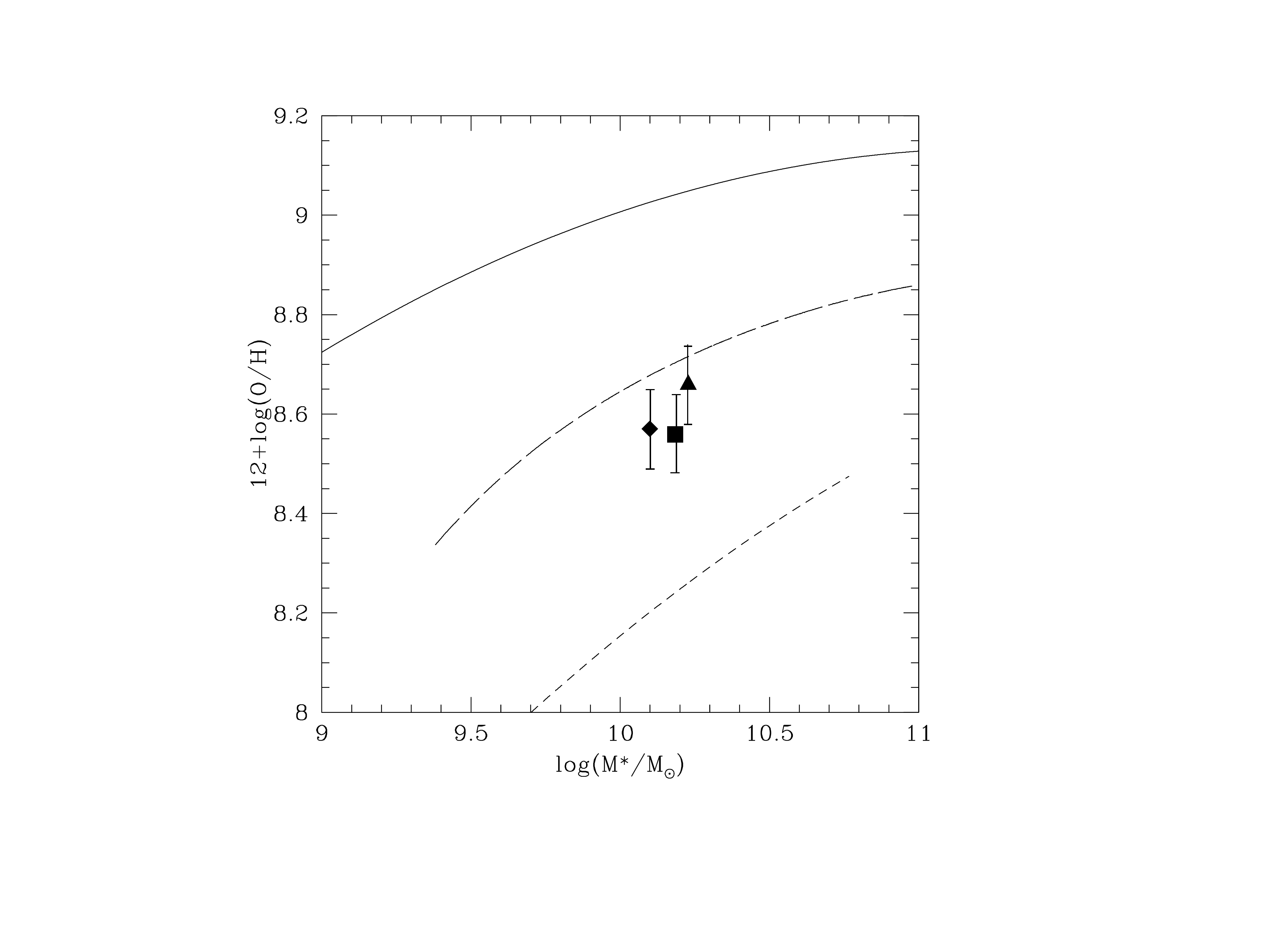}}
\end{center}
\caption{This is the stellar-mass--metallicity relation for the galaxies VVDS-3884 (diamond), VVDS-8666 (triangle) and VVDS-5183 (square). 
 The solid curve shows the relation at $z\sim0$ \citep{2004ApJ...613..898T}, the long-dashed curve shows the
relation at $z\sim2$ \citep{2006ApJ...644..813E}, the short-dashed curve the relation at $z\sim3$ \citep{2008A&A...488..463M}.}
\label{metal}
\end{figure}

Figure~\ref{metal} shows the position of these galaxies in the stellar mass-metallicity plane. 
In this figure, we also show the $z\sim0$ \citep{2004ApJ...613..898T}, $z\sim2$ \citep{2006ApJ...644..813E}, and
 $z\sim3$ \citep{2008A&A...488..463M} relations, the last two being rescaled to \citet{2004ApJ...613..898T} metallicities
 using equations from \citet{2008ApJ...681.1183K}.
They fall between the
$z\sim2$ and $z\sim3$ relations. We note however that the [O{\scriptsize~III}]/H$\beta$
\citep{2006ApJ...652..257L} calibration relies on the underlying relation between gas-phase oxygen abundance and ionization degree in the
parent sample. Thus, this calibration might overestimate metallicities for galaxies at high redshift, provided that they are likely
to show higher ionization states and higher [O{\scriptsize~III}]/H$\beta$ ratios for lower metallicities than the ones observed in the local universe.
Compared to
the local mass-metallicity relation, the galaxies of our sample show a $-0.47$ dex mean systematic shift in metallicity.

\begin{table*}
	\caption{Photometry}
	\resizebox{18cm}{!} {
	\begin{tabular}{cccccccccccccc}
		\hline
		Galaxy	&	$B_{AB}$ (1)	&	$V_{AB}$ (1)	&	$R_{AB}$ (1)	&	$I_{AB}$ (1)	&	$u_{AB}$ (2)	&	$g_{AB}$ (2)	&	$r_{AB}$ (2)	&	$i_{AB}$ (2)	&	$z_{AB}$ (2)	&	$J_{AB}$ (3)	&	$K_{AB}$ (3)	&	$3.6\,\mu$m (AB) (4)	&	$4.5\,\mu$m (AB) (4)	\\
		\hline
		VVDS-3884	&	24.65$\pm$0.09	&	23.77$\pm$0.06	&	23.83$\pm$0.06	&	23.39$\pm$0.08	&	---	&	---	&	---	&	---	&	---	&	24.04$\pm$0.31	&	23.29$\pm$0.20	&	22.0$\pm$0.1$^{\dagger}$	&	22.8$\pm$0.3	\\
		VVDS-8666	&	25.6$\pm$0.19	&	24.68$\pm$0.10	&	24.03$\pm$0.09	&	23.67$\pm$0.12	&	27.06$\pm$0.34	&	25.32$\pm$0.06	&	24.26$\pm$0.03	&	23.82$\pm$0.02	&	23.72$\pm$0.06	&	---	&	22.83$\pm$0.18	&	22.1$\pm$0.1	&	21.9$\pm$0.1	\\
		VVDS-7772	&	27.53$\pm$0.82	&	26.30$\pm$0.34	&	26.06$\pm$0.41	&	25.01$\pm$0.32	&	$>27$	&	26.31$\pm$0.19	&	25.74$\pm$0.13	&	25.52$\pm$0.13	&	$>27$	&	---	&	22.57$\pm$0.20$^{\ddagger}$	&	---	&	---	\\
		VVDS-5183	&	$>27$	&	25.37$\pm$0.24	&	23.95$\pm$0.07	&	23.68$\pm$0.10	&	---	&	26.17$\pm$0.22	&	24.15$\pm$0.04	&	23.77$\pm$0.03	&	23.45$\pm$0.08	&	---	&	---	&	22.7$\pm$0.2	&	22.8$\pm$0.3	\\	
		\hline 
	\end{tabular} 
}	
$^{\dagger}$ Confused source; $^{\ddagger}$\,The $K$-band magnitude is anomalously bright for VVDS-7772\\
 The columns are as follows:
(1) CFHT Legacy Survey observations
(2) TERAPIX CFHTLS-T0003 observations
(3)	SOFI/NTT observations
(4)	IRAC photometry 

\label{tab:photom}
\end{table*}
\begin{table*}
\caption{Stellar population properties}
\centering
\begin{minipage}{180mm}
	\begin{tabular}{cccccccccc}
		\hline
	 	Galaxy 	 	&	 	 $M_{V}$	 	&	 	 $M_{I}$ 	 	&	 	 $M_{\ast}$ (a)	 	&	 	 Age (b)	 	&	 	 	 	   $L_{UV}$ (c)  	 	&	 	  	     $SFR_{UV}$ (d)	 	&	 	 $SFR_{SED}$ (e) & $E(B-V)_{star}$ (f) \\ 
		\hline	
	 	VVDS-3884 	 	&	 	 $-22.40$ 	 	&	 	$-23.78$  	 	&	 	 $1.29$  	 	&	 	 $718$ 	 	&	 	 	 	    $23.85\pm1.38$ 	 	&	 	     	   $20\pm 1.1$ 	 	&	 	  $20.2$  	& $0$ \\  
	 	VVDS-8666 	 	&	 	 $-22.86$  	 	&	  $-23.68$	  	 	&	 	 $1.63$  	 	&	 	 $29$  	 	&	 		 	   $19.68\pm 1.60$ 	 	&	 	     	  $16\pm 1.3$ 	 	&	 	  $77.9$ &  $0.35$ 	\\ 
	 	VVDS-7772	 	&	 	 $-23.12^{\dagger}$ 	 	&	 	-- 	 	&	 	 $2.85^{\dagger}$	 	&	 $321^{\dagger}$	 	 	&	 		 	$3.09\pm1.17$	 	&	 	 	$2.6\pm0.97$	 	&	 	$0^{\dagger}$  &	$0^{\dagger}$  \\ 
	 	 VVDS-5183 	 	&	 	  -- 	 	&	 	$-23.19$  	 	&	 	 $1.51$ 	 	&	 	 $4.79$ 	 	&	 	 	 	    $30.63\pm 0.85$ 	 	&	 	     	  $25\pm0.7$ 	 	&	 	 $532$  & $0.40$	\\		\hline	
	\end{tabular}\\  
$^{\dagger}$\,The $K$-band magnitude, from which $M_V$ is derived, is anomalously bright for VVDS-7772. We use a post-starburst SED with no dust to fit it, rather than continuous star formation.\\
 The columns are as follows:
(a) Stellar Mass ($10^{10} M_{\odot}$)
(b) Stellar population age (Myr) from SED fitting,
(c) Luminosity ($ \times  10^{+28}$ \ergshz) from rest-frame UV continuum flux,
(d) Star formation rate ( $M_{\odot}$ yr$^{-1}$) estimated from rest-frame UV continuum flux,
(e) Star formation rate ( $M_{\odot}$ yr$^{-1}$) estimated from SED fitting.
(f) Dust reddening derived from the SED fitting.
\end{minipage}
\label{sfr_mass}
\end{table*}
\subsection{Photometry, SED modelling and Stellar population properties }
\label{sec:seds_masses}
We use broad-band photometry from the optical to the mid-infrared to determine the spectral energy distribution, and through fitting spectral
evolutionary synthesis models we constrain the masses and ages of the stellar populations, and derive an independent estimate of the star rate
to compare with that from the rest-frame optical nebular lines. We use the photometry from the VVDS catalog of the VVDS Deep Field
0226-04\footnote{Available from http://cencosw.oamp.fr/VVDSphot/VVDS/vvds.html} \citep{2003A&A...410...17M,2004A&A...417..839L} in the $BVRI$
filters, obtained with the CFH12K camera on CFHT. For some galaxies this is supplemented by $J$ and $K_S$ imaging with SOFI on the ESO NTT
\citep{2005A&A...442..423I,2008A&A...482...81T}. We supplement these magnitudes with the recent CFHT Legacy Survey observations of this area
obtained with MegaCam, using photometry in the $griz$ from the TERAPIX CFHTLS-T0003 release\footnote{http://cencosw.oamp.fr/CFHTLS/ presents
the CFHT-LS magnitudes and photometric redshifts} \citep{2006A&A...457..841I}. We note that all magnitudes used here are on the AB system. For
the VVDS optical/near-IR photometry we are using MAG\_AUTO (that is, ``total" magnitudes derived by SExtractor).
\begin{figure*}
\begin{center}
\resizebox{0.5\columnwidth}{!}{\includegraphics{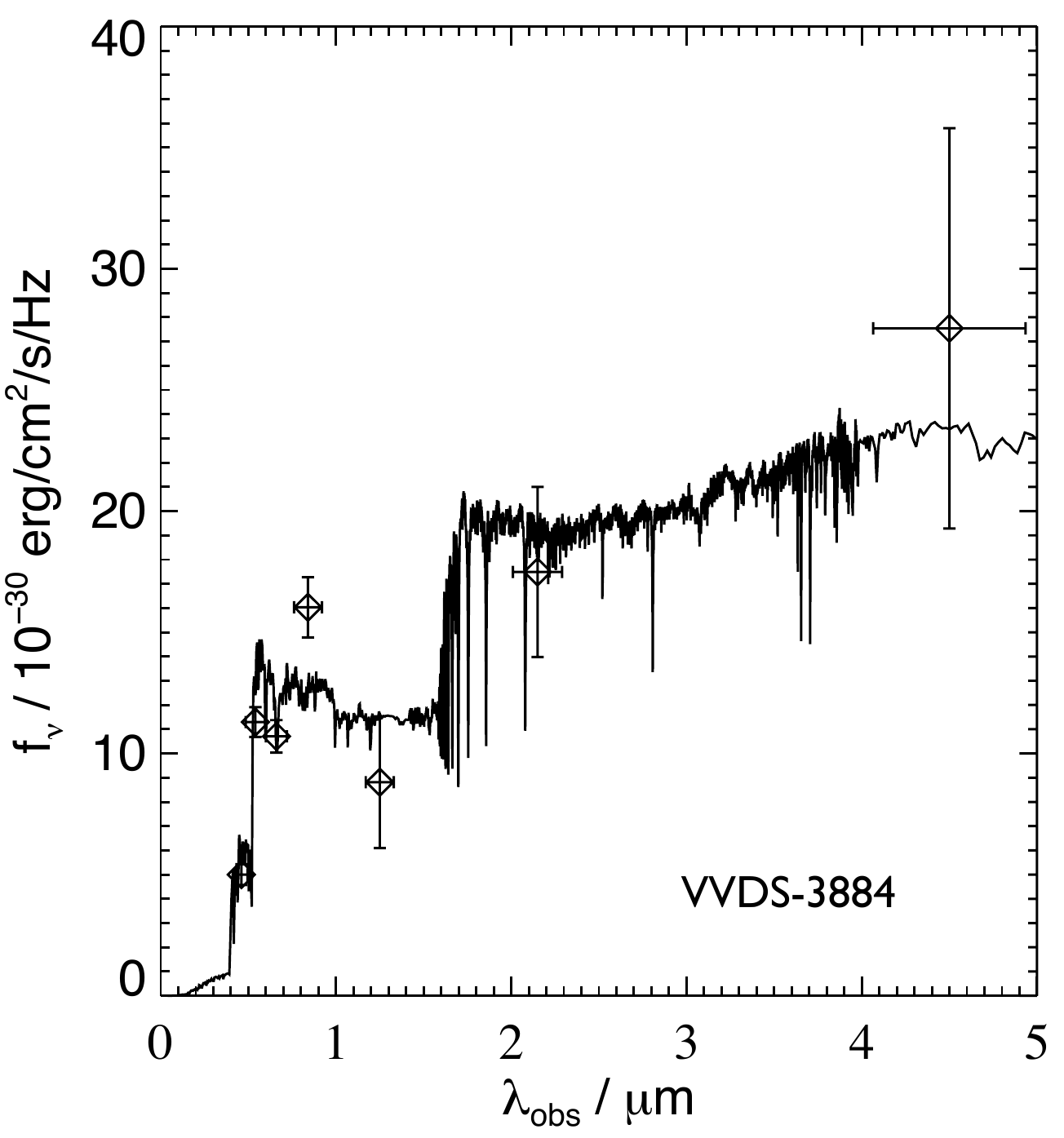}}
\resizebox{0.5\columnwidth}{!}{\includegraphics{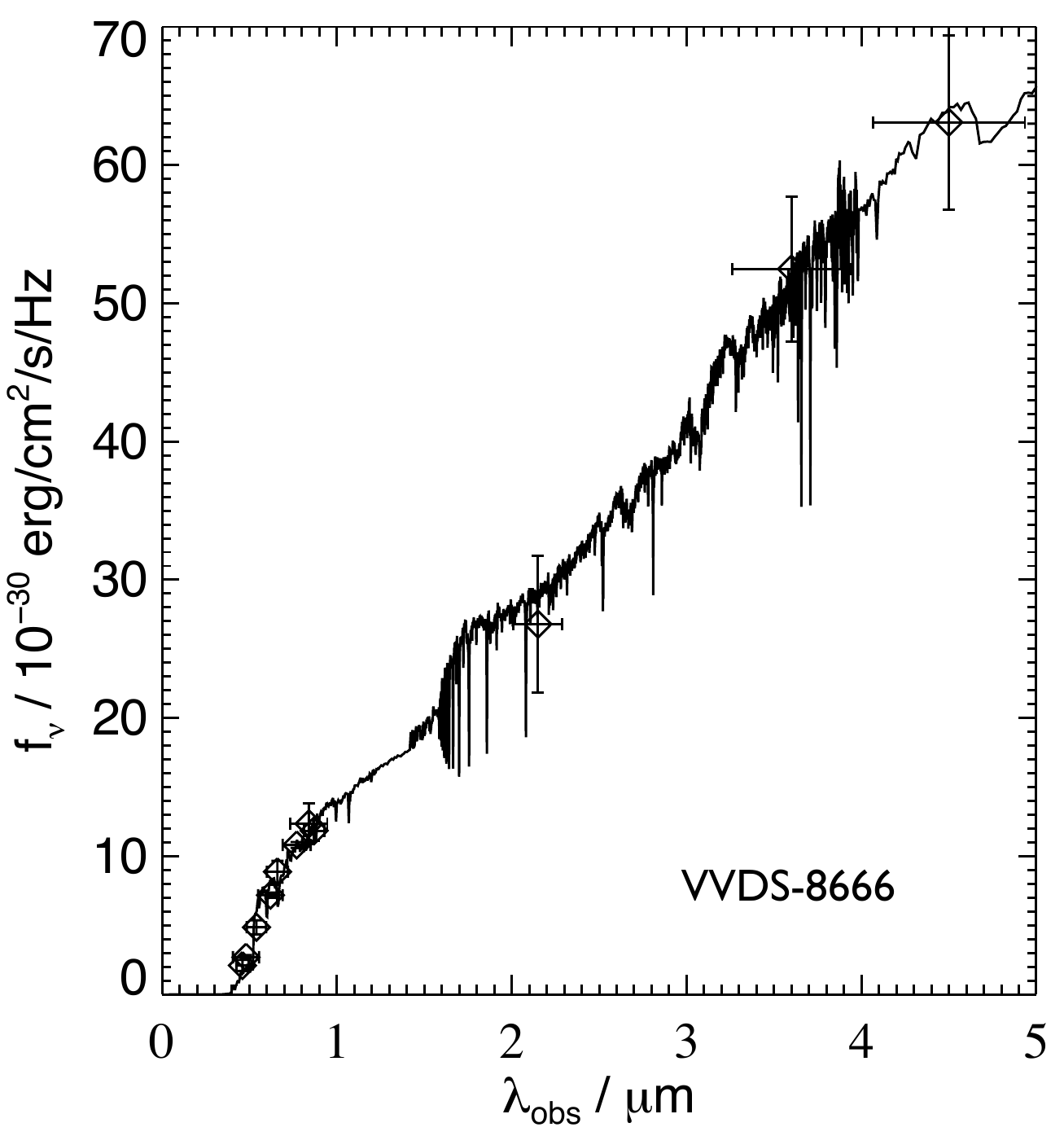}}
\resizebox{0.505\columnwidth}{!}{\includegraphics{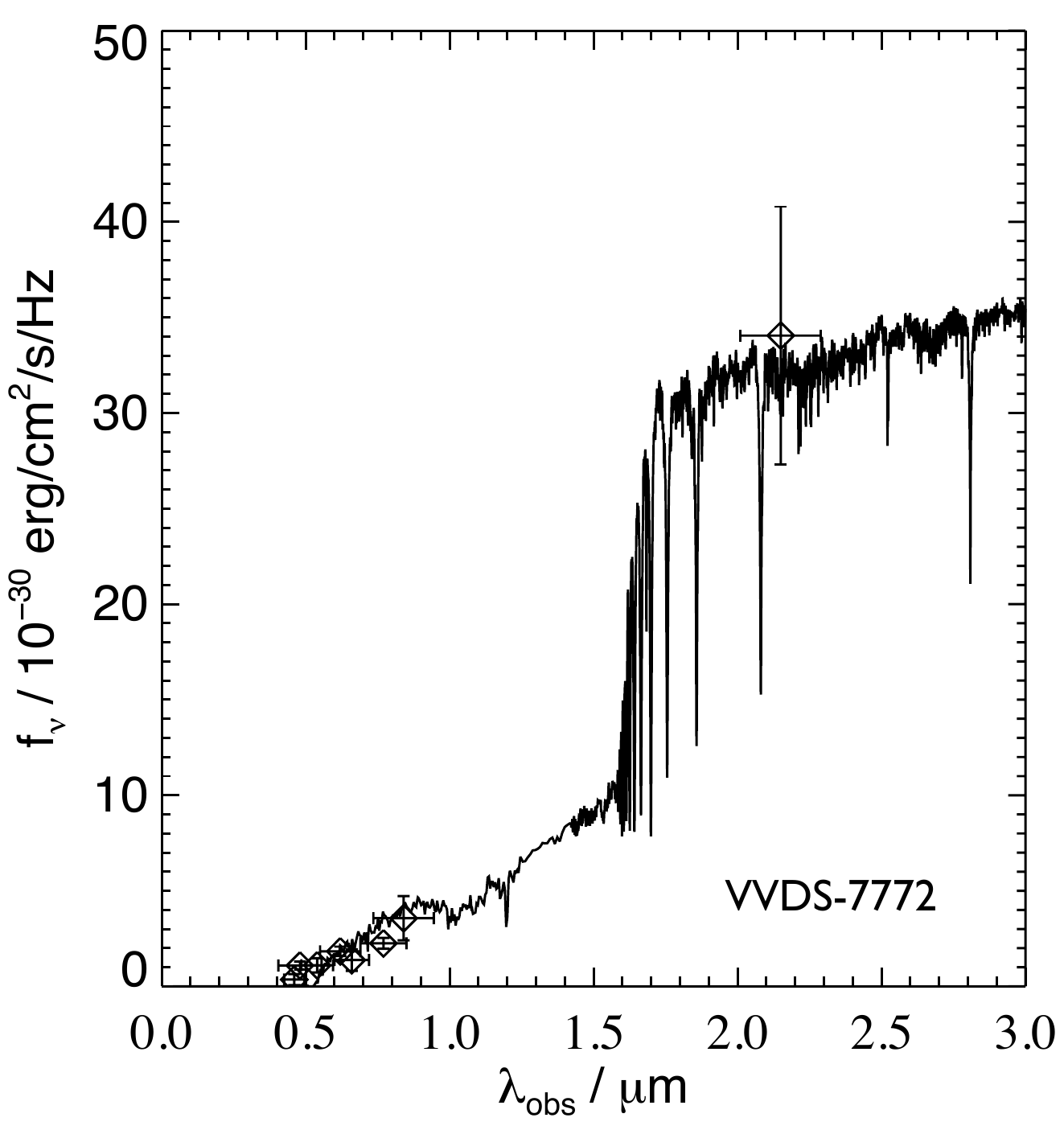}}
\resizebox{0.5\columnwidth}{!}{\includegraphics{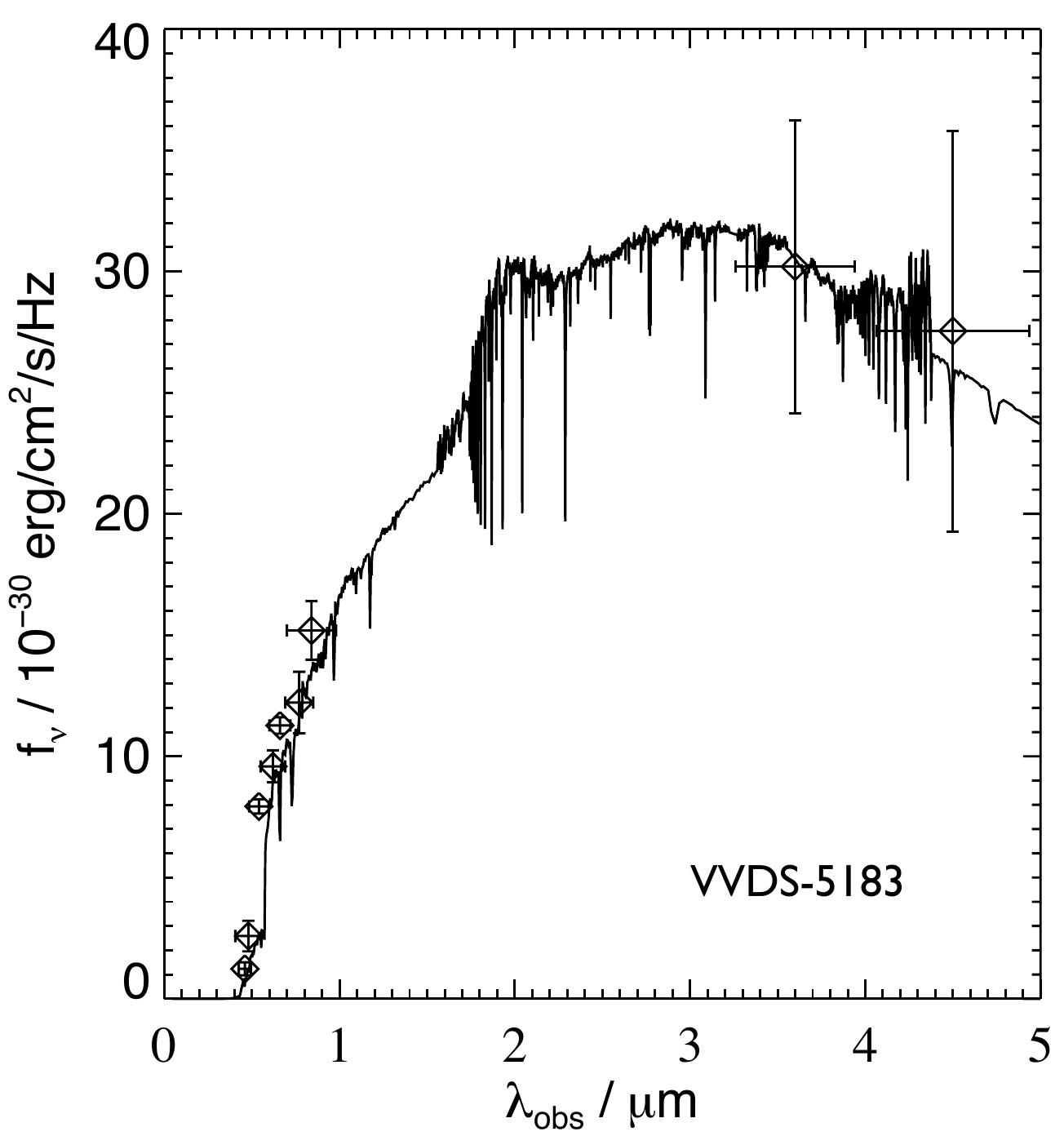}}
\end{center}
\caption{The spectral energy distribution for VVDS-3884, VVDS-8666, VVDS-7772 and VVDS-5183 from the broad-band CFHT optical/near-IR
photometry, and Spitzer/IRAC channels 1 \& 2. The best-fit continuous star formation model
is overlaid. Bruzual \& Charlot (2003) models have been
used with a \citet{2003PASP..115..763C} IMF and solar metallicity.}
\label{SED_666}
\end{figure*}
%
\begin{table*}
\caption{Nebular emission properties}
\resizebox{18cm}{!} {
\begin{tabular}{ccccccccccc}
\hline
Galaxy 	&	$A_{neb}$ (a)	&	 $\Sigma_{gas}$ (b)	&	$M_{gas}$ (c)	&	$\mu$ (d)	&	 AGN? (e) 	&	 12+log(O/H) (f) 	&	 $L(\mathrm{H}\beta$) (g) 	&	 $SFR_{neb}^0$ (h) 	&	$E(B-V)_{gas}$ (i) & $SFR_{neb}$ (j)\\
\hline
VVDS-3884 	&	78.6	&	521	&	4.1	&	0.76	&	 $\approx40\%$ chances 	&	 $8.57\pm0.02$ 	&	 $542\pm27$ 	&	 $125\pm4$ 	&	 $0$ & $125$\\
VVDS-8666 	&	70.2	&	471	&	3.3	&	0.67	&	 no 	&	 $8.66\pm0.02$ 	&	 $429\pm19$ 	&	 $97\pm19$	&	 $0.35$ & $427$\\
VVDS-7772 	&	37.4	&	$<$90	&	$<$0.1	&	$<$0.16		&	 \emph{yes} 	&	 - 	&	 $<29$ (2$\sigma$) 	&	 $<5.1$ (2$\sigma$) 	&	$0$ & $<5.1$\\
VVDS-5183 	&	74.6	&	838	&	6.3	&	0.81	&	 $\approx40\%$ chances 	&	 $8.56\pm0.02$ 	&	 $949\pm38$ 	&	 $231\pm57$ 	&	 $0.40$ & $1257$\\
\hline
\label{nebular_sfr}
\end{tabular}
}
The columns are as follows: (a) area of the \oiiib\ emission in kpc$^2$, 
(b) gas surface density $M_{\odot}$ yr$^{-1}$ pc$^{-2}$,
(c) Gas mass ($10^{10} M_{\odot}$),
(d) Gas fraction $\mu = M_{\rm gas}/(M_{\rm gas} + M_{\ast})$,
(e) is the nebular spectrum dominated
by an AGN (`no' implies it is dominated by star formation) 
(f) gas-phase oxygen abundance,
(g) Luminosity ($\times10^{40}\mbox{\ensuremath{\,}erg\ensuremath{\,}s}^{-1}$),
(h) star formation rate ($M_{\odot}\,\mbox{yr}^{-1}$)  estimated from the H$\beta$ luminosity
(i) reddening suffered by gas, derived from SED fitting
(j) dereddened star formation rate.
\end{table*}

There is also {\em Spitzer Space Telescope} imaging of this field (the XMM LSS), obtained under the Spitzer Wide-area Infrared Extragalactic
Survey (SWIRE, \citealt{2005AAS...207.6301S}). We measured the IRAC photometry from version 4 of the SWIRE team's reduced data
products\footnote{Available from http://data.spitzer.caltech.edu/popular/swire/20061222\_enhanced/XMM\_LSS/irac/ }. In the reduced dataset, the
IRAC pixels have been rebinned by a factor of 2 during the ``drizzling" process from their original $1\farcs2\,{\rm pix}^{-1}$, so the pixel
scale is now $0\farcs6$. We convert the units of the mosaics from MJy/sr (the calibration from the Spitzer PBCD pipeline) to $\mu$Jy, requiring
multiplying by 8.74. As with the optical/near-IR imaging, we work in AB magnitudes, where 1\,$\mu$Jy corresponds to 23.93 AB mags. Images in
all four Spitzer channels we available (with central wavelengths $3.6\,\mu$m, $4.5\,\mu$m, $5.8\,\mu$m \& $8.0\,\mu$m), although the less
sensitive channels 3 and 4 had no detections in the vicinity of our SINFONI observations, with $3\,\sigma$ limits of $m_{AB}<21.3$ and $21.1$
at $5.8\,\mu$m \& $8.0\,\mu$m respectively. For channels 1, 2, 3 \& 4 we used circular apertures of radius 2, 2, 2.5 \& 3 pixels ($2\farcs4$,
$2\farcs4$, $3\farcs$ \& $3\farcs6$ diameter) respectively, and an aperture correction of 0.7\,mag (see \citealt{2005MNRAS.364..443E}).
Once magnitudes in each of the different wavebands had been obtained, the photometric data were then used to construct SEDs for each of our
selected sources. We made use of the \citet{2003MNRAS.344.1000B} isochrone synthesis code (hereafter B\&C), utilising the Padova-1994
evolutionary tracks (preferred by B\&C). The models span a range of 221 age steps approximately logarithmically spaced, from $10^5$\,yr to
$2\times 10^{10}$\,yr, although here we discount solutions older than $\sim 2\times 10^9$\,yr (the age of the Universe at $z\approx 3$). The B\&C
models have 6900 wavelength steps, with high resolution (FWHM 3\,\AA ) and 1\,\AA\ pixels over the wavelength range 3300\,\AA\ to 9500\,\AA\,
and unevenly spaced outside this range. 
Throughout this paper, we adopt the \citet{2003PASP..115..763C} initial mass function (IMF).
From the range of possible star formation histories (SFH) available, we
focus on a constant star formation rate (SFR), as these Lyman-break galaxies are selected on their rest-UV continuum (i.e. have ongoing star
formation). This star formation rate history is also used in the work of \citet{2006ApJ...647..128E}. For the constant SFR model, the B\&C
template normalization is an SFR of $1\,M_{\odot}\,{\rm yr}^{-1}$. We also consider the possibility that the optical--infrared colours of
objects within our sample could be due to intrinsic dust reddening, rather than an age-sensitive spectral break. We adopted the empirical
reddening model of \citet{1997AJ....113..162C}, suitable for starburst galaxies. 
Table \ref{sfr_mass} summarise the properties  of the stellar population (absolute magnitudes, stellar mass, population age, extinction, etc.) obtained for the galaxies of our sample.
\subsection{Star Formation rates} 
We computed star formation rates for the four galaxies in our sample deduced from the H$\beta$, the SED fitting and from the rest-frame UV continuum emission. 

Ultraviolet-derived star formation rates were calculated from the broadband optical photometry, using $I$-band for VVDS-5183 at $z=3.7$
and $R$-band for the other galaxies at $z\approx 3.3$ to determine the UV continuum level around 1500\,\AA . In the absence of dust, the UV
continuum density per unit frequency ($f_{\nu}$) is approximately flat between the Lyman-$\alpha$ and the Balmer break for a constant star
formation rate. The mean luminosity in frequency units is calculated using the following equation:
\begin{equation}
L_{1500} \:\:($\ergshz$ ) = 10^{(-0.4\cdot m(AB))} \times 3631\cdot10^{-23} \times 4\pi \frac{D_{L(cm)}^2}{1+z}
\end{equation}
 We can then  deduce $SFR_{UV}$ following \citet{k98} scaled to a  \citet{2003PASP..115..763C} IMF: 
\begin{equation}
\textup{SFR } (\msun \textup{ yr}^{-1}) = 0.83 \times 10^{-28}\;L_{1500} \:\:($\ergshz$ )
\end{equation}
and the star formation rates derived from the rest-frame UV are shown in Table~\ref{sfr_mass}.

We also derived star formation rates from the H$\beta$ emission-line luminosity. Given that no other Balmer line is observed, we are not able
to compute the dust attenuation from the standard Balmer-decrement method. Thus, we decide to use instead the dust attenuation determined from
the SED fitting. Following \citet{2006ApJ...647..128E}, we adopt  $E_{star}(B-V)=E_{gas}(B-V)$. \citet{2009A&A...495..759A} have shown that when the dust attenuation is not computed from the Balmer-decrement
method (e.g. with the observed H$\alpha$/H$\beta$ ratio), the standard scaling-law which relates the emission-line luminosities 
and the star formation rates, i.e. $SFR=a \times L(line)$ \citep{k98}, is not valid anymore. It has to be replaced instead by a
power-law relation, i.e. $SFR=a \times L(line)^b$ with $b \neq 1$.  We used the \citet{2000ApJ...533..682C} extinction law.
We thus combine equations 6 and 14 of \citet{2009A&A...495..759A} to estimate the nebular star formation rates of our
galaxies. We multiplied the results for star formation rate by a factor of $0.88$ to convert from \citet{2001ASPC..228..187K} IMF to
\citet{2003PASP..115..763C} IMF. 
The resulting dust-corrected star formation rates from H$\beta$ are shown in Table~\ref{nebular_sfr}.
 We found high values for  $SFR_{neb}$ corrected which should be taken with caution. It is indeed difficult to estimate a reliable dust extinction from SED fitting at such high redshifts, 
where the relations between dust, age, and metallicity might be very different to what is observed and relatively well understood at low redshift. 
Thus, this values are indicative, and we use the non-dereddened values of the star formation rate, $SFR_{neb}^0$, in the rest of the paper.

\subsection{Gas Masses and Gas Fractions} 
We computed the gas mass using the empirical global Schmidt law, the correlation between star formation surface density and gas density (see
\citealt{1998ApJ...498..541K}). Assuming the galaxies in our sample follow such a law, we used the $SFR_{neb}^0$ and the spatial extent of the
\oiiib\ emission, an area $A_{neb}$ (measured from the 2D intensity map of the \oiiib\ emission, and deconvolved with the PSF, see \ref{morpho_prop}),
to compute the star formation surface density (as is done in \citealt{2006ApJ...647..128E}). From the star formation surface density, we deduced
the gas density using equation (4) in \citet{1998ApJ...498..541K}. We then deduced the gas masses by $M_{\rm gas}=\Sigma_{\rm gas} A_{neb}$.
Using the stellar masses determined from the SED fitting, we can derive the gas fraction by $\mu = M_{\rm gas}/(M_{\rm gas} + M_{\ast})$.
Table~\ref{nebular_sfr} summarizes the results obtained for the four galaxies of our sample. All of these objects,  except  VVDS-7772, have high values of the gas
fraction ($\mu > 0.5$), similar to that found in the $z\approx 2-2.5$ sample of \citet{2009ApJ...697.2057L}.
\section{Properties from spatially resolved kinematic} 
\label{kine_prop} 
\begin{figure*}
\begin{center}
\resizebox{2.0\columnwidth}{!}{\includegraphics{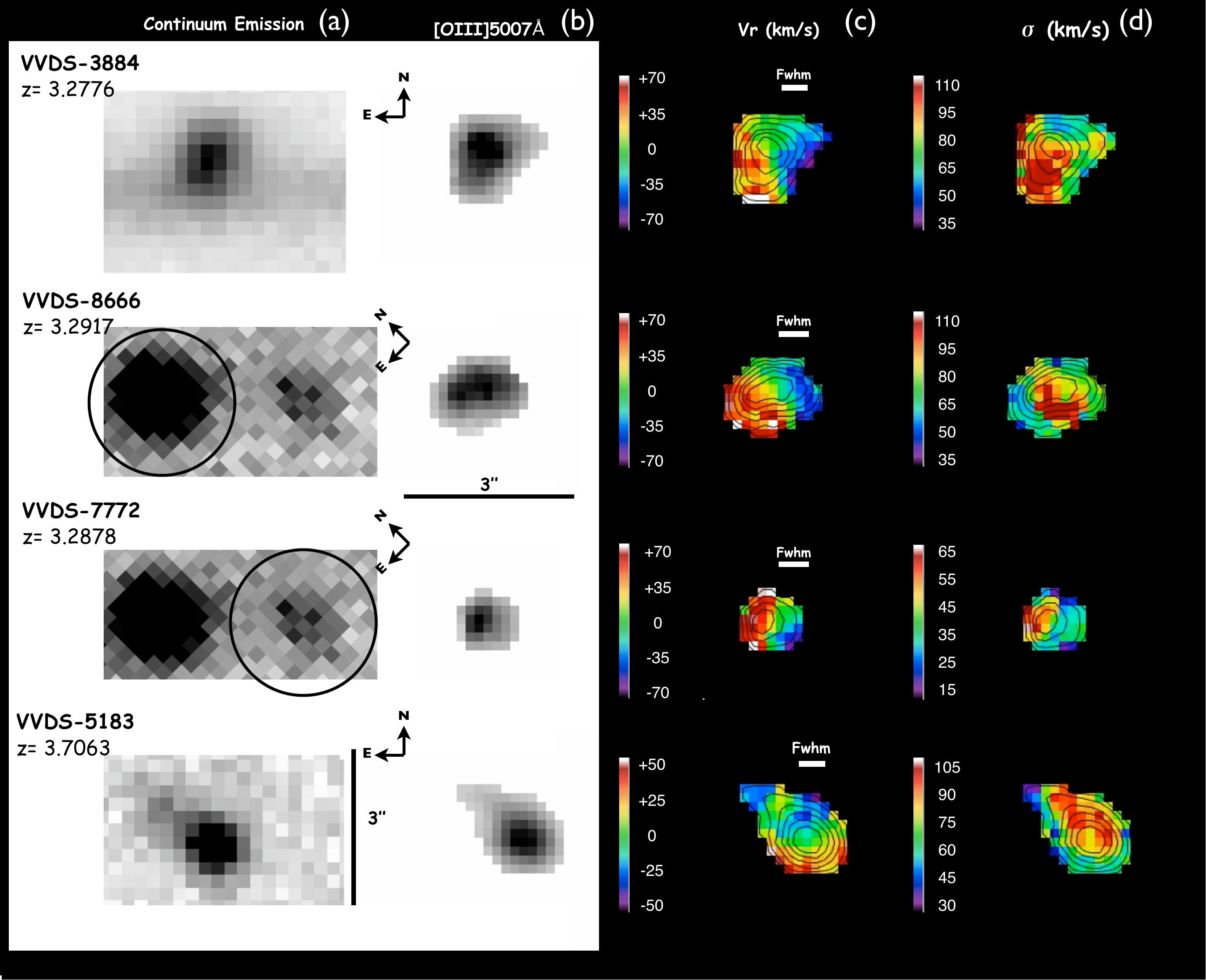}}
\end{center}
\caption{From left to right: (a) CFHTLS image ($r$-band for VVDS-3884, VVDS-8666 and VVDS-7772; $i$-band for VVDS-5183), (b) \oiiib\ flux map, (c) \oiiib\ velocity field, and (d) \oiiib\ velocity dispersion map obtained from Gaussian fits to the SINFONI data cubes after smoothing spatially with a two-dimensional Gaussian of FWHM $=$ 3 pixels. The CFHTLS images and \oiiib\ flux maps are color-coded with a linear scaling such that the values increase from light to dark. These data have been acquired with the 125$ \times $250mas sampling configuration of SINFONI.  An angular scale bar of 3\arcsec, the redshift computed from the \oiiib\ emission line and the spatial rsolution are indicated for each galaxy.}.
\label{map1}
\end{figure*}
\subsection{Kinematic measurements} 
We produced maps of the dynamics for the galaxies in our sample by using E3D, the Euro3D visualization tool \citep{2004AN....325..171S}, and
code for the fitting and analysis of kinematics (e.g., \citealt{2004ApJ...615..156S,2005A&A...429L..21S}). Our main goal was to determine the
kinematics of the ionized gas using the strongest emission line, \oiiib\ . For each spatial pixel (spaxel) we fit the \oiiib\ emission line
region to a single Gaussian function, in order to characterize the emission line, and a pedestal to characterize any spectral continuum. We
first smoothed the reduced cubes spatially with a Gaussian of $\rm FWHM = 3~pixels$ ($0\farcs 37$). The line flux, FWHM, central wavelength and
the continuum (pedestal) were then fitted. From the results of this fitting we obtained maps (see Fig.~\ref{map1}) of the \oiiib\ emission line
intensity, the relative radial velocity map ($V_{r}$) and the velocity dispersion ($\sigma$). The dispersion map was corrected for the
contribution of the instrumental dispersion, as determined from the FWHM of unblended and unresolved sky lines. Error maps for the velocity and
dispersion measurements were also computed, which are dominated by the effect of random noise in fitting the line profile, which produces large
errors at low S/N. Errors ranged from $~6$ km s$^{-1}$ to $~24$ km s$^{-1}$ for the maps.
\subsection{Morphologic properties}
\label{morpho_prop} 
The left-most panel, (a), of Figure \ref{map1} shows the UV rest-frame morphology around 1500\,\AA , with panel (b) showing the maps of the
ionized gas morphology for the galaxies in our sample. For the four objects of the sample, the UV rest-frame morphology and the morphology of
the ionized gas appear similar. VVDS-3884 presents a bright concentrated central region, with a diameter of 2.3 kpc. Two fainter
regions slightly elongated toward the South and the Northwest surround the bright nucleus. VVDS-8666 exhibits an elliptical morphology
with a large peak in intensity at its centre in both the $R$-band image and the line map. The galaxy VVDS-7772 looks compact, showing a
peak in the ionized gas map slightly  offset to the North-East. VVDS-5183 consists of a bright nucleus and a low surface brightness
elongated region extended toward the North-East.

For VVDS-8666 we used the GALFIT software \citep{2002AJ....124..266P} on the CFHT $R$-band image, to deduce the morphological parameters such as
the center, the position angle, and the axial ratio. Assuming that the rest-UV light is dominated by stars in an inclined disk, we derived a
position angle of $\theta = 41^{\circ}$ East of North and an inclination of $i = 51^{\circ}$ (where $i=0^{\circ}$ would be an edge-on disk).
For the other galaxies in our sample, the images were too faint, compromised by a stellar diffraction spike, or lacked good nearby PSF stars
for GALFIT to return satisfactory fits.

We assume that the rest-UV morphology of the young stellar population will closely trace that of the nebular emission from the ionized gas, and
hence we infer the morphological parameters of the galaxies from the [O{\scriptsize~III}] map. We fit a 2D gaussian with the FWHM along the major axis, with the
axial ratio and position angle as free parameters. A first guess of the position angle is estimated from the velocity map as the direction
where the gradient in the velocity profile is maximum. It is then allowed to vary between $\pm15^{\circ}$ in the rotation modelling (see
\ref{model}) which finally gives the best-fit position angle. If the galaxy is a disk, then the inclination is estimated as $\cos^{-1}(b/a)$
where $a$ is the radius of the major-axis (along the first guess of the position angle) and $b$ the radius of the minor-axis (at $90^{\circ}$).
We measured the area of the nebular emission (deconvolved with the PSF) from the two-dimensional \oiiib\ flux map for each of the galaxies
(Table \ref{nebular_sfr}).
\subsection{Rotation modelling}
\label{model}
Each galaxy is modelled by a pure, infinitely thin, rotating disk. The parameters of a model are the kinematic centre ($x_{0}$,$y_{0}$), the
position angle (PA), the inclination ($i$), the velocity offset ($V_{\mathrm{s}}$) of the centre relative to the integrated spectrum, and the
velocity curve $V_{\mathrm{c}}(r)$ where $r$ is the radius from the kinematic centre. We also use the true physical velocity dispersion
$\sigma_{0}$ as a model parameter, which we assume is constant, and represents the thickness of the rotating disk. The radial velocity $V$ for
any point is then defined with standard projection equations. Note that the position angle gives the direction of positive radial velocities.
The velocity offset accounts for redshift uncertainties. We define the velocity curve by two parameters, following \citet{2007ApJ...658...78W}: the maximum velocity $V_{\mathrm{max}}$ and the radius $r_{\mathrm{c}}$ where this maximum velocity is achieved. In this model, below
$r_{\mathrm{c}}$ the velocity is defined as
\begin{equation}
V_{\mathrm{c}}(r) = \frac{r}{r_{\mathrm{c}}}\times V_{\mathrm{max}}\
\end{equation}
The model computes
$V_{rot}$, the asymptotic maximum rotation velocity at the plateau of the rotation curve, corrected for inclination ($V_{rot}=V_{max}/\sin i$).

In reality, the spatial resolution is limited by the seeing and the spaxel size. The observed radial velocity is thus the weighted convolution
of the true radial velocity by the point spread function (PSF). The weights come from the flux map of the line used to compute the velocity
map: a spaxel where the observed line flux is negligible will not contribute to the convolution. Additionally, the observed velocity dispersion
accounts not only for the true physical dispersion, but also for the variations in the velocity field inside the width of the PSF. We generate
a map of the velocity dispersion where the velocity gradient (determined from the modelling) has been subtracted. This yields a better measure
of the intrinsic local velocity dispersion, rather than a raw value of $\sigma$ which may be inflated if the velocity gradient across a spaxel
is large. From this map, we compute the flux-weighted mean velocity dispersion, $\sigma_{mean}$ (Table~\ref{kinematic_properties}). We also
measure a velocity width for the emission lines from the extracted 1D spectrum (i.e. the total integrated light); this $\sigma_{1D}$ includes
contributes both from the random motions and any ordered rotation or bulk motions.

The beam smearing introduced by the PSF causes two significant effects. First, the observed velocity map appears smoothed, so the velocity
gradient and the maximum velocity are underestimated. Second, the observed dispersion map shows a peak of dispersion near to the kinematic
centre.

We can reproduce modelled velocity and dispersion maps by applying mathematically the same weighted convolution to the ideal velocity field.
These modelled maps are compared to the observed maps by a $\chi^{2}$ minimization. The parameters which minimize the $\chi^{2}$ are computed
in two successive grids with increasing resolution. First guesses for the kinematic centre, the position angle, and the inclination are set
from the morphology.
\begin{figure*}
\begin{center}
\resizebox{1.2\columnwidth}{!}{\includegraphics{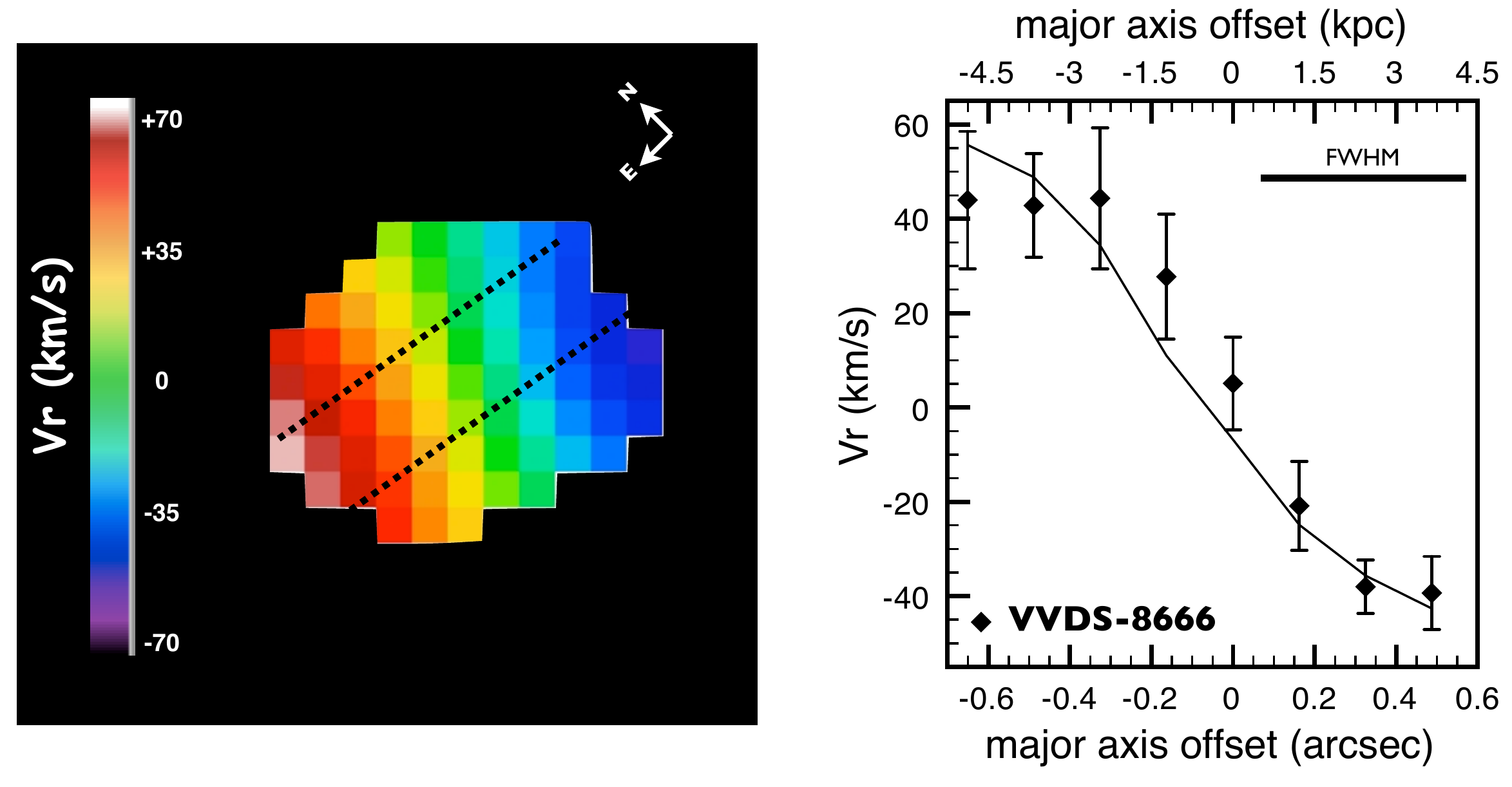}}
\end{center}
\caption{Two-dimensional disk  model after convolution with a 0.58\arcsec\ PSF for VVDS-8666  with the same spatial location as the observed velocity map  (left). 1D rotation curve (right; full diamond) extracted using an `idealized' slit from the SINFONI \oiiib\ observed velocity map (see map (c) in Fig.~\ref{map1}), overlaid with the best-fit model (solid line).}  
\label{comp2D_666}
\end{figure*}
\begin{figure*}
\begin{center}
\resizebox{1.2\columnwidth}{!}{\includegraphics{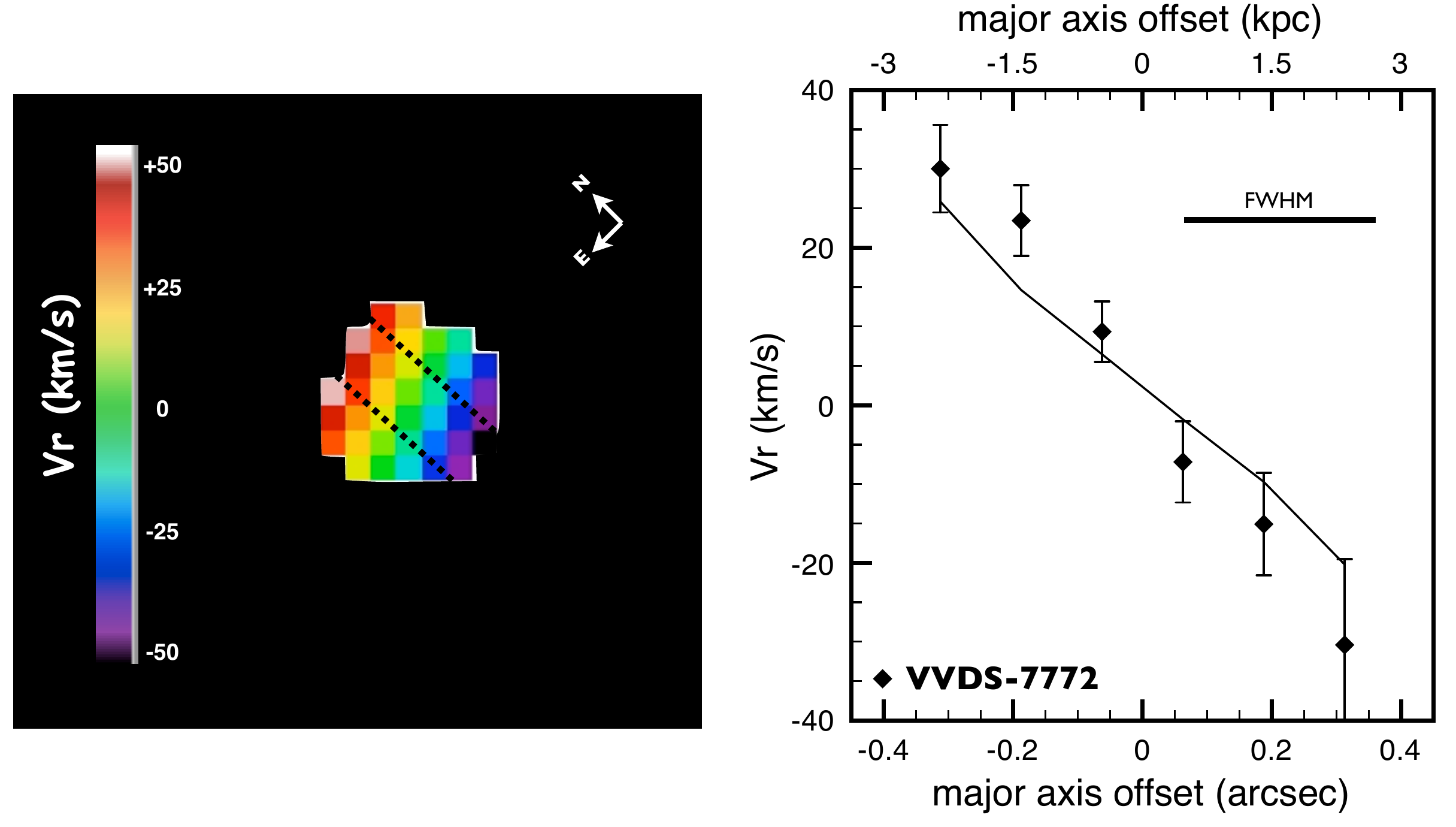}}
\end{center}
\caption{Maps and curve are the same as in Fig.~\ref{comp2D_666} but for the galaxy VVDS-7772. }  
\label{comp2D_772}
\end{figure*}
\subsection{Properties of the ionized gas kinematics} 
\subsubsection{VVDS-8666}
The VVDS-8666 galaxy displays a well-resolved velocity gradient over $\sim5$\,kpc in projected distance with a peak-to-peak amplitude of
92\,km\,s$^{-1}$ (uncorrected for inclination). There is some evidence of a flattening of the rotation curve, particularly in the Eastern-most
region of the galaxy. Taking into account the fact that these data are seeing limited, the smearing effect of the seeing decreases the
peak-to-peak observed amplitude while mostly keeping the overall shape of the velocity shear. This shear appears to be aligned with the
morphological major axis defined by the ionized gas, and is also nearly spatially coincident with the peak in the dispersion map. This,
combined with the fact that the nebular line flux distribution is centrally concentrated, leads us to conclude that the observed shear is
consistent with a rotating gaseous disk rather than a merger. Figure \ref{comp2D_666} shows the \oiiib\ velocity map recovered from the
rotation modelling (see Figure \ref{map1}, panel c) and convolved with the PSF. It show also the one-dimensional relative velocity curves along
the kinematic major axis for both the SINFONI observed velocity shear and the best-fit model. The velocity is well matched by our simple
rotating disk model, leading to an inclination-corrected $v_{rot} \sim 91 \pm 26$ km\,s$^{-1}$. However, the ratio of the rotation velocity to
the velocity dispersion is large ($v_{rot}/\sigma_{mean} = 1.52$) compared with lower-redshift samples of galactic disks
($v_{rot}/\sigma_{mean} =10-20$, \citealt{2006ApJ...638..797D}) which probably indicates that this object is not mainly supported by rotation,
and there are significant random motions. The presence of random motion is in particular confirmed by a peak of $\sim50$ km/s, located to the
north of the galaxy, in the residuals of the velocity maps after subtraction of the rotating model.
\subsubsection{VVDS-7772}
The velocity map of VVDS-7772 (the companion galaxy of VVDS-8666) shows a smoothly varying gradient along the North-South axis, but
without evidence for a flattening. However, this galaxy is very compact and only marginally resolved. We note that the velocity fields of both
VVDS-7772 and VVDS-8666 are aligned along their displacement vector and close to the major axis of the resolved galaxy
VVDS-8666, so this combined system might plausibly be part of a single rotating disk. The best fit simple rotation model give an
asymptotic velocity of $V_{rot} \sim 98 \pm 45$ km/s which is reached in $ \sim 5.4$\,kpc radius. The $\sigma$-map of the ionized gas of
VVDS-7772 shows strong variation with position, as can be seen in panel (d) of Fig. \ref{map1}, which prevents us from modelling the true
physical dispersion; the weak constraint is $\sigma_0=0\pm28$\,km\,s$^{-1}$.  In particular, $\sigma$ peaks around 58\,km\,s$^{-1}$ in the
south-western edge of the galaxy, slightly offset with respect to the flux peak for which $\sigma$ decreases to around 50\,km\,s$^{-1}$, and
drops further to $\sim$\,30\,km s$^{-1}$ in the faint emission of the north-eastern region. However it is uncertain whether the peak of $\sigma
\sim$\,58\,km\,s$^{-1}$ represents a real feature or is simply noise in the extreme edges of the galaxy. Due to the compact size of this objects
($r \sim 3$kpc), it is difficult to draw conclusions on the nature of the kinematics from our seeing-limited observations.
\begin{figure*}
\begin{center}
\resizebox{1.2\columnwidth}{!}{\includegraphics{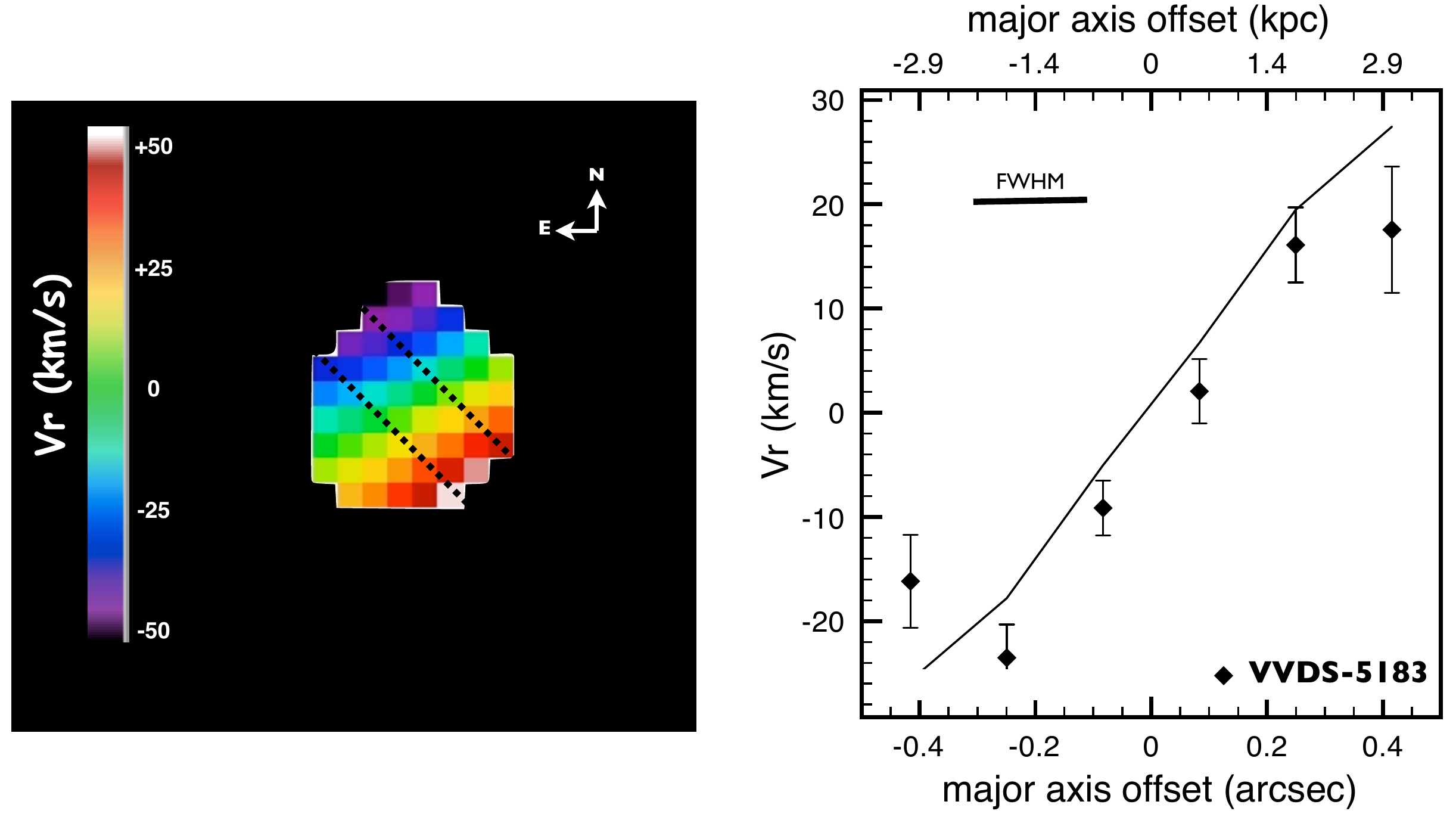}}
\end{center}
\caption{Maps and curve are the same as in Fig.~\ref{comp2D_666} but for the  galaxy VVDS-5183.}  
\label{comp2D_183}
\end{figure*}
\subsubsection{VVDS-3884}
This galaxy presents an irregular velocity field, which is inconsistent with a  smoothly varying velocity gradient along the major axis.
It was therefore impossible to fit this disturbed shear with a single rotation model. The $\sigma$-map shows a peak in the local velocity dispersion  displaced from the centre of the galaxy, again inconsistent with disk rotation. This galaxy has the largest velocity dispersion in our sample ($\sigma \sim$\,126\,km\,s$^{-1}$). 
\subsubsection{VVDS-5183}
This galaxy is dominated by an extended emission line component at the systemic redshift with a faint secondary extension located $\sim$ 2 kpc
to the North-East. The two spatially distinct components are more apparent in the data cube (i. e. the velocity map) than in the \oiiib\
intensity maps. Between the two components, the velocity seems to return to the systemic (central) velocity of the main component, i.e.\ the
velocity of the whole system is not monotonically increasing with position. The faint secondary feature may represent a kinematically distinct
star-forming region or a small galaxy in the process of merging with the brighter system. The primary component exhibits a smoothly varying
velocity field well fit by a rotating disk aligned with the morphological major axis. However the maximum of the velocity shear is small
($V_{max}\sim 30$\,km\,s$^{-1}$), and once again this galaxy has significant random motion (high velocity dispersion) with $v_{rot}/\sigma_{mean} = 0.95$. 
\begin{table*}
\caption{Kinematics properties}
\label{kinematic_properties}
\resizebox{18cm}{!} {
	\begin{tabular}{ccccccccccc}
		\hline
ID	&	i (1)	&	PA (1)	&	$\sigma_{mean}$ (2)	&	$\sigma_{1D}$ (3)	&	$V_{shear}$	(4) &	$V_{rot}$ (5) &	$r_{gas}$ &	$r_{model}$ (6) &	$M_{dyn}(V_{rot}) / 10^{10}M_{\odot}$ (7) &	$M_{dyn}(\sigma_{1D}) / 10^{10}M_{\odot}$ (8)\\
	\hline
 VVDS-3884	&	90$\pm$40	&	--	&	72$\pm$29	&	126$\pm$9	&	50$\pm$12	&	--	&	5.27$\pm$0.6	&	-- &	--	&	65.7 \\
 VVDS-8666	&	51$\pm$25	&	41$\pm$25	&	60$\pm$16	&	79$\pm$6	&	46$\pm$16	&	91$\pm$26	&	4.89$\pm$0.6	&	0.4$\pm$3&	2.9	&	23.8 \\
VVDS-7772 	&	35$\pm$18	&	73$\pm$51	&	0$\pm$28	&	53$\pm$12	&	33$\pm$17	&	98$\pm$45	&	3.00$\pm$0.9	&	1.4$\pm$2 &	12.1 ($>0.76^{\ddagger}$) 	&	6.5 \\
 VVDS-5183	&	50$\pm31^{\dagger}$	&	$-10\pm 30^{\dagger}$	&	78$\pm 11^{\dagger}$	&	95$\pm$3	&	28$\pm$6	&	74$\pm$42	&	5.84$\pm$0.4 (3.58$^{\dagger}$)	&	4.3$\pm 3^{\dagger}$ &	$21.6^{\dagger}$ ($>0.65^{\ddagger}$) 	&	41.3 \\
		\hline
	\end{tabular}  
}
$^{\dagger}$\,estimated from the primary component only.\\
$^{\ddagger}$\,where we do not observe a flattening of the rotation curve, a robust lower-limit on the $M_{dyn}$ is given by $V_{shear}^2\,r_{gas}\,/\,G$\\
 The columns are as follows:
(1) morphological parameters: inclination `i' and position angle `PA'.
(2) Flux-weighted mean velocity dispersions deduced from the rotation modelling,
(3) Velocity dispersion of integrated spectrum from the \oiiib fwhm.
(4) Shear velocity $v_{\rm shear} = \frac{1}{2} (v_{\rm max} - v_{\rm min})$ measured on the \oiiib\ velocity map,
(5) asymptotic velocity inferred from the rotation modelling
(6) radius inferred from the rotation modelling for which  $v_{\rm shear} = V_{rot}$
(7) The total dynamical mass inferred from the velocity dispersion of the integrated spectrum
(8) The total dynamical mass inferred from the rotation modelling
\end{table*}
\subsection{Dynamical masses} 
 From the nebular line kinematics determined from our 3D data cubes, we are able to estimate the dynamical masses of the systems, subject to
several caveats. Firstly, we assume that the nebular emission from the gas traces Keplerian motions (i.e. the gas is not outflowing/inflowing,
and is in orbit rather than unvirialized as might be the case during a merger). Secondly, we cannot say with certainty that any of our galaxies
are traditional rotationally-supported gaseous disks -- the ground-based seeing precludes accurate measurement of morphological parameters
(such as the crucial inclination angle). Thirdly, the measured velocity dispersion is significant when compared with any velocity gradient
across these galaxies, implying that a putative disk may not be supported purely by rotation.
 
 We have measurements of the spatially varying velocity field ($v$) and velocity dispersion ($\sigma$) from our 3D data cubes, and in the 1D
extracted spectrum we have a flux-weighted measure of the overall velocity spread in the galaxy ($\sigma_{1D}$) which arises from the combined
effects of any systematic velocity shift across the galaxy and random motions.
We can use this  $\sigma_{1D}$ as a crude estimate of the dynamical mass using the formula:
\begin{equation}
M_{dyn}(\sigma_{1D}) = \frac{C \sigma_{\rm 1D}^2 r_{gas}}{G}
\label{dynmass.eqn}
\end{equation}
The factor $C$ depends on the geometry of the system (in particularly the density profile): for a uniform sphere, $C=5$, and Erb et al.\ (2006c) derive $C=3.4$ for the more realistic scenario of a gas-rich disk with an average inclination angle. Using this value of $C=3.4$, in  Table~\ref{kinematic_properties} we present dynamical masses inferred from the $\sigma_{1D}$ and the spatial extent of the \oiiib\ line emission in our data cubes ($r_{gas}$) measured along the major axis and deconvolved with the seeing.

 Of course, the availability of 3D data cubes means we can model any spatially-resolved systemic velocity shift, as might arise in a rotating
disk (see Section~\ref{model}). For objects where the velocity shear is well fitted by the simple rotation model, we compute the dynamical
masses using the formula: 
\begin{equation} 
	M_{dyn}(V_{rot}) = \frac{V_{rot}^2\,r_{model}}{G} \label{dynmassrot.eqn} 
\end{equation} 
	where $V_{rot}$ (the turnover of the rotation curve) has been inclination-corrected (a factor of $\sin i$, fitted by the model from an initial
estimate based on the observed ellipticity). Both this asymptotic velocity, and the radius of the turnover ($r_{model}$) are inferred from our
model fits to the observed velocity maps to correct for the the effect of beam-smearing (due to the seeing). Only in one case (VVDS-8666)
do we convincingly see this flattening of the rotation curve within the radius probed by our detected \oiiib\ emission (Fig.~\ref{comp2D_666}),
from which we infer $M_{dyn}=2.9\times 10^{10}\,M_{\odot}$. For galaxies VVDS-7772 and the primary component of VVDS-5183, the
inferred turnover from the models lies beyond the range of data, and hence must be treated as unreliable (see Fig.~\ref{comp2D_772} and
Fig.~\ref{comp2D_183} respectively); we can place a lower limit on the dynamical mass from the measured maximum velocity and spatial extent of
the nebular emission: $M_{dyn}>V_{shear}^2\,r_{gas}\,/\,{G}$, which corresponds to $M_{dyn}>7\times 10^{9}\,M_{\odot}$ in both these cases. The
galaxy VVDS-3884 does not demonstrate evidence of significant ordered rotation, and we were unable to fit a simple rotating disk model in
this case.
\begin{figure}
\begin{center}
\resizebox{0.9\columnwidth}{!}{\includegraphics{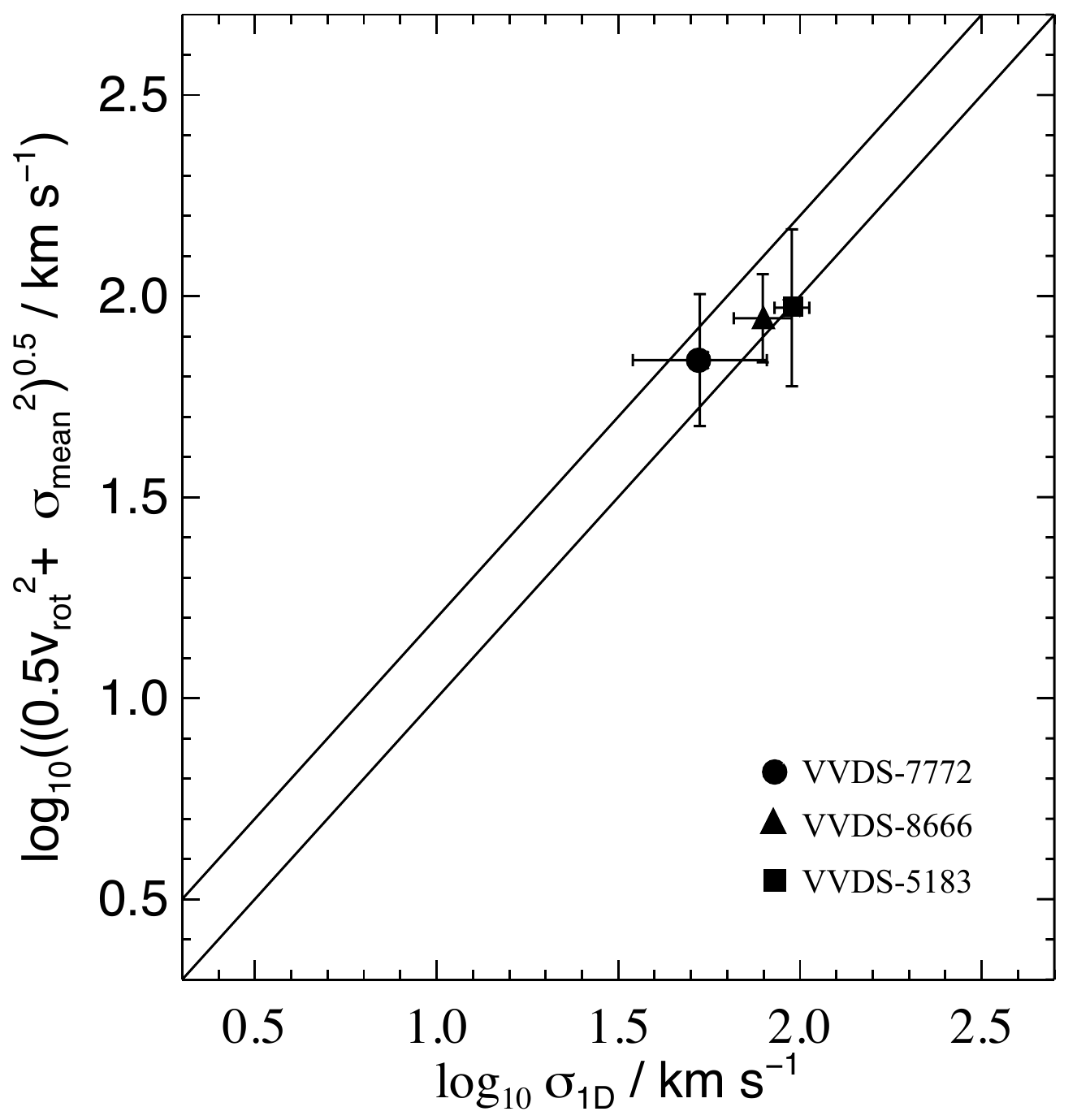}}
\end{center}
\caption{Combined velocity, $\whalf^2 = 0.5 \vrot^2 + \sigtwod^2$ versus integrated linewidth \sigoned. The diagonal lines are
the same as in \citet{2006ApJ...653.1027W}, i.e.\ a $1:1$ line and the \citet{1997MNRAS.285..779R} $\sigma = 0.6 V_c$ line ($V_c$ is the
circular velocity). Line width and \whalf\ are correlated; the 0.5 pre-factor makes the combined velocity width a better estimate
of velocity dispersion, so that the correlation is tighter and the galaxies closer to the $1:1$ line. Our galaxies show good agreement with
this.}
\label{Weiner2006_fig19}
\end{figure}
\section{Discussion and Conclusions} 
\label{summary} 
We present the results of a study of the kinematic properties of the ionized gas in four star-forming galaxies at redshift $z \sim 3$, from the
spatially-resolved spectra of the [O{\scriptsize~III}] \& H$\beta$ nebular emission lines. We found that the ionized gas in these objects has high velocity
dispersions ($\sigma_{mean}\approx 60-70$\,km\,s$^{-1}$). We have also found that all our galaxies (except VVDS-7772) contain large
quantities of gas as compared to their stellar mass and their dynamical mass inferred from rotation. With their high SFRs, it is likely
that these objects are undergoing episodes of strong star formation. Due to our seeing-limited observations, it is difficult to classify these
objects with confidence as either `disks' or mergers. In the case VVDS-8666 and its companion (VVDS-7772), we classify them as a ``heated disk system" since for both galaxies the velocity
structure of the ionized gas appears to be consistent with rotation. In fact it is unlikely that
these objects have a dynamically cold rotating disk of ionized gas, due to the significant velocity dispersions. VVDS-8666 has the highest
metallicity, probably indicating a later evolutionary stage.
The galaxy VVDS-3884 possesses a particularly high nebular star formation rate as compared to its SED star formation rate which
is consistent with this object experiencing a major recent burst of formation.
 This burst might originate from a major merging event, consistent with its complex kinematics. Finally, VVDS-5183 is mostly consistent with a merger on to a rotating disk. It has a huge star formation
rate, a large fraction of gas and a low metallicity.

Due to the uncertainties on the inclination angle for the objects in our sample, it is not possible to investigate the traditional Tully-Fisher
relation using \vrot. However, we can give an insight into a key question: Can we use the line widths of integrated nebular emission to get a
good estimation of the dynamical masses of galaxies showing low values of $v/\sigma$ ? Figure \ref{Weiner2006_fig19} plots the combined velocity
scale \whalf\ against the one-dimensional line width \sigoned\ for the three galaxies for which we found a consistency with the presence of
rotation. The combined velocity scale $\whalf^2 = 0.5 \vrot^2 + \sigtwod^2$ is estimated from the asymptotic velocity  ($V_{rot}$) and
the flux-weighted mean velocity dispersion, both inferred from the rotation modelling (see Figure \ref{Weiner2006_fig19}). It represents an estimator,
corrected from the smearing effect of the seeing  (see section \ref{model}), of rotation velocity together with the presence of random motions. \whalf\ is also well known
to be strongly correlated with \sigoned\ \citep{2006ApJ...653.1027W,2007ApJ...660L..35K}. We found a very good correlation between \whalf\ and
\sigoned\ for the VVDS-8666, VVDS-7772 and the primary component of VVDS-5183. This shows not only that the one-dimensional line
width \sigoned\ can be used as a kinematic measure when estimating dynamical masses, but that \vrot\ alone cannot suffice to trace the
majority of the dynamical mass. In fact, using only \vrot\ to investigate Tully-Fisher relation will push galaxies with $\sigtwod>\vrot$ to
erroneously low dynamical masses (see also \citealt{2006ApJ...653.1027W}).
\begin{figure}
\begin{center}
\resizebox{0.9\columnwidth}{!}{\includegraphics{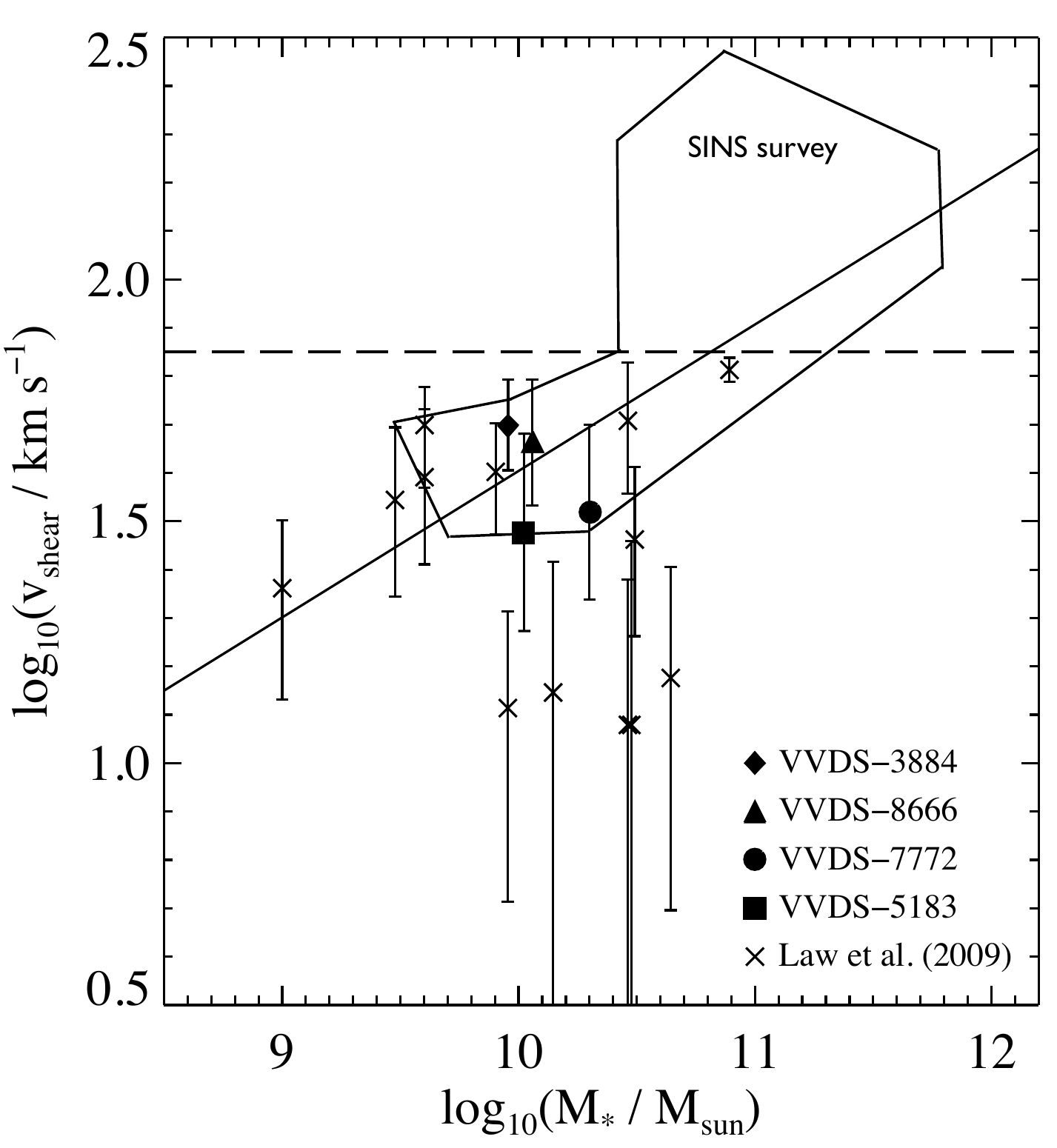}}
\end{center}
\caption{Maximum line-of-sight velocity shear vs. stellar masses for our four galaxies. The dashed line denotes $\langle$log$(\sigma_{\rm mean})\rangle$ of the \citet{2009ApJ...697.2057L} galaxies sample. We found that all our galaxies fall below this line. The solid line indicates the fit to the \citet{2009ApJ...697.2057L} sample (log ($v_{\rm shear}$/km$s^{-1}$)$ = 0.30 \times$ log($M_{\ast}/M_{\odot}$) - 1.39). The box on this figure denotes the region of `SINS' galaxy population (see text).}
\label{Law2009_fig9}
\end{figure}

Despite the small size of our sample, we can try to compare our results with those found by the `SINS' survey
\citep{2006ApJ...645.1062F,2006Natur.442..786G,2007ApJ...671..303B,2008ApJ...682..231S,2008ApJ...687...59G}, also using SINFONI on VLT, and the
\citet{2009ApJ...697.2057L} study using OSIRIS on Keck. Our four galaxies have a mean assembled stellar mass of $\langle \log (M_{\ast}/M_{\odot})
\rangle = 10.4$, while  the `SINS' galaxy population has a larger mass of $\langle \log (M_{\ast}/M_{\odot}) \rangle = 10.9 - 11.0$. More comparable to our sample are the 13 galaxies detected by \citet{2009ApJ...697.2057L}, selected by the BX technique, with stellar masses of $\langle \log
(M_{\ast}/M_{\odot}) \rangle \approx 10.1$. We also found, in agreement with \citet{2009ApJ...697.2057L}, that such galaxies of intermediate mass at
$z \sim 3$ possess extremely large quantities of gas in comparison to their stellar mass. The typical $v/\sigma$ for our sample is around $\sim 0.4 -
1.5$. The \citet{2009ApJ...697.2057L} sample have $\sim 0.8$ while the `SINS' galaxies have a range of $\sim 2 - 4$.
 In contrast to \citet{2009ApJ...697.2057L}, we have found evidence of AGNs in our sample. The connection between AGN and star formation could play an important role in shaping galaxies at $z \sim 3-4$.

Figure \ref{Law2009_fig9} plots the maximum observed shear velocity of galaxies, $v_{\rm shear}$, as a function of stellar mass. 
Our $z=3-3.7$ objects appear to have values of  $v_{\rm shear}$ similar to the highest values detected by \citet{2009ApJ...697.2057L} but 
corresponding to the lowest values displayed by the galaxies of the `SINS' survey. In fact at similar stellar masses our galaxies tend to have higher values of $v_{\rm shear}$   than  \citet{2009ApJ...697.2057L} sample but similar values of $v_{\rm shear}$ to the \citet{2006ApJ...645.1062F} sample.
We also found in general that the highest mass galaxies do not tend to have the highest velocity shear among our objects, similar to the \citet{2009ApJ...697.2057L} sample.
Therefore, our sample at $z= 3-3.7$ seems to comprise intermediate mass and intermediate shear velocity objects,  in contrast to 
the sub-sample of \citet{2009ApJ...697.2057L} which seems to have negligible velocity shear and the more massive objects, and the `SINS' survey sample which have appreciable old populations where stable rotation dominates.

The similarity in properties of our sample of galaxies at $z \sim 3-3.7$  with that of \citet{2009ApJ...697.2057L} at $z \sim 2-2.5$, seems to confirm that these high-redshift objects have not yet turned a significant fraction of their gas into a sizeable stellar population. Such galaxies also tend to possess high local velocity dispersions due to random motions intrinsic to the gas. If any velocity shear is present, it is not likely to be significant compared with the random motions, and probably occurs in a non-stable disk system. These galaxies also exhibit strong episodes of star formation and already possess metal rich gas.

Drawing general conclusions about the nature and the properties of the dynamical structure of distant galaxies based on the
current set of observations is obviously very challenging, given the small size of current samples and the spatial resolution limitations. However, initial insights into the kinematic properties of $z \sim 2-4$ galaxies on scales of a few kiloparsecs  have been revealed by the recent IFU studies, although larger statistical samples and unbiased selection
criteria are needed to give a quantitative picture of the mechanisms at play in such objects. It already appears that the dynamical state of
galaxies during this early period, where the star formation appears to be intense, cannot be easily classified based on what is observed at low redshift. Specifically, the star-forming galaxies at $z\sim 3$ (even those with measurable velocity gradients) exhibit large velocity
dispersions quite unlike the cold rotationally-supported disks seen at lower redshift.
\section*{Acknowledgments} 
We would like to thank Aprajita Verma and Matthias Tecza for helpful discussions. We are very grateful to the VLT Observatory for accepting this programme. The authors also thank Markus Hartung for help obtaining the observations during the observing runs  and Sebastian S\'anchez for providing the code to create the kinematics maps from the 3D cube.  The anonymous referee is greatly acknowledged for providing useful and constructive comments. The authors wish to recognize and acknowledge the significant contribution of the VVDS collaboration in providing the targets. In particular we thank Thierry Contini who helped greatly with the target selection.
Part of this work was supported by the Marie Curie Research Training Network {\it Euro3D; contract No.  HPRN-CT-2002-00305}.  M. Lemoine-Busserolle is supported by the Gemini Observatory, which is operated by the Association of Universities for Research in Astronomy, Inc., on behalf of the international Gemini partnership of Argentina, Australia, Brazil, Canada, Chile, the United Kingdom, and the United States of America.

\label{lastpage}

\end{document}